\documentclass[a4paper,11pt]{article}
\pdfoutput=1 
\usepackage{jheppub} 
\usepackage[T1]{fontenc} 
\usepackage{lmodern}

\usepackage[utf8]{inputenc}
\usepackage{amsmath,amsfonts,amssymb,amsthm}
\usepackage{graphicx}
\usepackage{contour}
\usepackage{subcaption}
\usepackage{microtype}
\usepackage{tikz-feynman}
\definecolor{light-gray}{gray}{0.95}
\def\be{\begin{equation}}
\def\ee{\end{equation}}
\def\ba{\begin{eqnarray}}
\def\ea{\end{eqnarray}}

\preprint{%
\begin{minipage}{3cm}
 QMUL-PH-19-01
\end{minipage}
}

\title{Eikonal Scattering in Kaluza-Klein Gravity}

\author{Arnau Koemans Collado}
\author{and Steven Thomas}
\affiliation{Centre for Research in String Theory, School of Physics and Astronomy,\\Queen Mary University of London,\\Mile End Road, E1 4NS London, United Kingdom}

\emailAdd{a.r.koemanscollado@qmul.ac.uk}
\emailAdd{s.thomas@qmul.ac.uk}

\abstract{We study eikonal scattering in the context of Kaluza-Klein theory by considering a massless scalar field coupled to Einstein's gravity in 5D compactified to 4D  on a manifold $M_4\times S^1 $. We also examine various different kinematic limits of the resulting eikonal. In the ultra-relativistic scattering case we find correspondence with the time delay calculated for a massless particle moving in a compactified version of the Aichelburg-Sexl shock-wave geometry. In the case of a massless Kaluza-Klein scalar scattering off a heavy Kaluza-Klein mode a similar calculation yields the deflection angle of a massless particle in the geometry of an extremal, $Q=2M$, Einstein-Maxwell-dilaton 4D black hole. We also discuss a related case in the scattering of dilatons off a large stack of $D0$-branes or $D6$-branes in dimensionally reduced $D=10$ type IIA string theory.}

\begin{document} 
\maketitle
\flushbottom
\section{Introduction }
\label{intro}
It has long been understood that in considering the process of 2 $ \rightarrow $ 2 elastic scattering of particles coupled to gravity, the high energy or Regge limit is dominated by the exchange of highest spin massless particles, namely the graviton. As one considers processes involving more and more graviton exchanges the series of ladder diagrams produced sum to an exponential phase which is usually referred to as the eikonal \cite{'tHooft:1987rb,Muzinich:1987in,Kabat:1992tb,Giddings:2010pp,Akhoury:2013yua,Melville:2013qca}. This behaviour has been explicitly shown to be the case for massive scalars coupled to Einstein gravity as well as to theories including higher derivative corrections to the Einstein-Hilbert action \cite{Camanho:2014apa}. These phenomena have also been observed in supergravity and superstring theories involving more exotic field content where both elastic and inelastic 2 $ \rightarrow $ 2 processes are seen \cite{Amati:1987wq,D'Appollonio:2010ae,D'Appollonio:2013hja,Collado:2018isu}. There is also the question of subleading corrections (in the high energy limit) to the eikonal which correspond to diagrams which for example include 3-point interactions between the gravitons themselves. These corrections also exponentiate into their own series, the subleading eikonal \cite{Akhoury:2013yua,Collado:2018isu}.

In perhaps the simplest case of a single massless real scalar coupled to Einstein gravity, the leading and subleading eikonal have been computed assuming 2 $ \rightarrow $ 2 scattering in a non-compact background manifold ${\mathbb R}^{1,D-1} $ with $D-1$ non-compact spatial dimensions. In this paper we want to focus on the case when one of the spatial dimensions is compactified on a circle of radius $R$, i.e. with a background  manifold,  ${\mathbb R}^{1,D-2} \times S^1$. There is a choice of where to orient the $S^1$ with respect to the plane of scattering. Here we assume it is along one of the transverse direction, so that  the transverse momentum exchange $q = (q', n/R) $ where $q $ and $ q' $  are continuous momenta  in $ {\mathbb R}^{1,D-1}  $ and ${\mathbb R}^{1,D-2}$ receptively and $n/R$, $n  \in {\mathbb Z } $  is the quantized momenta along the $S^1 $.  This simple  Kaluza-Klein compactification  produces a surprisingly rich theory in  ${\mathbb R}^{1,D-2}$  where we find an infinite Kaluza-Klein tower of charged states emerging from the scalar and graviton in ${\mathbb R}^{1,D-1}$. In particular from this lower dimensional viewpoint, 2 $ \rightarrow $ 2 scattering of scalar particles now involves generally massive and charged (with respect to the $U(1) $ gauge field emerging from the metric in higher dimension) Kaluza-Klein particles, involving  both elastic and inelastic processes. The latter involve massive Kaluza-Klein scalars which change their species via exchange of a massive spin-2, spin-1 and spin-0 states in the Kaluza-Klein tower. By contrast elastic scattering of scalars involves only the exchange of a massless graviton, photon and dilaton.  

Our first aim is to prove that in the high energy limit for 2 $ \rightarrow $ 2 scattering of scalars in the compactified model, the leading order one-loop contribution to the amplitude is proportional to the square of the  leading tree-level contribution when written in impact parameter space.  This provides strong evidence that the series should exponentiate if we consider higher loop contributions and allows us to identify the eikonal phase in the compact case. This calculation is carried out assuming general Kaluza-Klein masses  $m_i^2 = n_i^2/R^2$, $i =1 \ldots 4 $ for the 4 scalars involved (with the restriction that $\sum_i n_i = 0$ implied by momentum conservation along $S^1$). As such it involves both elastic and inelastic processes.\par
We then specifically focus on the elastic scattering of Kaluza-Klein scalars and analyse the eikonal in various kinematic limits. These should be related to the deflection angle derived in the corresponding background metrics. For example, in the non-compact case in $D=4$, it is well known \cite{Camanho:2014apa} that a for a model with  two  real scalars of mass $m_1, m_2$, the eikonal in the ultra-relativistic limit $s \gg |t|$, $s \gg m_1^2, m_2^2$  is related to the time delay calculated in the background of a Aichelburg-Sexl shock wave metric. On the other hand, the  limit where one of the scalars is light and the other ultra heavy,  e.g. $m_1 =0 $ and $m_2^2 > s \gg |t| $, the corresponding eikonal is related to the deflection angle in the background of a Schwarzschild black hole of mass $m_2 $ \cite{Akhoury:2013yua,Bjerrum-Bohr:2016hpa,Collado:2018isu,Bjerrum-Bohr:2018xdl}.

We revisit these limits for our compactified model, focussing mainly on the case where we compactify from $D=5$ to $D=4$. In the ultra-relativistic limit, for fixed Kaluza-Klein masses $m_1$, $m_2$  namely  $s' \gg |t'|$, $s' \gg m_1^2 , m_2^2$, we find the eikonal phase is related to a compactified version of the Aichelburg-Sexl shock wave metric, which has appeared in the previous  literature in the study of shock waves in brane world scenarios \cite{Emparan:2001ce}. In the second kinematic  limit we consider elastic scattering of a massless Kaluza-Klein scalar off a heavy Kaluza-Klein scalar of mass $ m_2 $ with   $m_2^2 > s' \gg |t'| $,  we find the leading eikonal is related to the leading order contribution (in inverse powers of the impact parameter $b$) of the deflection angle in a Schwarzschild black hole of mass $m_2$. However this is not the whole story. In fact at leading order in $1/b$,  the same result would hold for a charged dilatonic black hole background which is a solution of Einstein-Maxwell-dilaton (EMd) theory \cite{Horne:1992zy}. These are precisely the black hole backgrounds we should expect to be relevant in our model because the heavy Kaluza-Klein scalars are electrically charged and also couple to the dilaton field. This becomes clear when we move on to consider the subleading contributions to the eikonal which a priori contribute terms at order $1/b^2 $ to the deflection angle. These corrections  involve  the exchange of massless  $U(1)$ gauge fields and dilatons as well as gravitons between the two scalars. We find that because of the precise charge mass relation $Q=2M$ (in appropriate units) for the Kaluza-Klein states, these subleading corrections to the eikonal vanish. In terms of the corresponding deflection angle we understand the vanishing of the subleading $1/b^2 $ terms as a consequence of the extremal $Q=2M$ limit of the deflection angle in the background of a EMd black hole.

This paper is structured as follows. In section \ref{kktheory} we introduce the action for Einstein gravity  coupled to a real massless scale in $D$ dimensions and its compactification to $D-1$ dimensions  on a circle, resulting in a Kaluza-Klein tower of charged massive scalars coupled to the massless graviton, dilaton and $U(1)$ gauge field. Although there are also the massive Kaluza-Klein excitations of states of these aforementioned fields, they contribute to inelastic scattering of the  massive charged scalars in $D-1$ dimensions, we will focus on the eikonal of elastic scattering processes in the present paper. In section \ref{bornapprox} we remind the reader of the known result of 2 $\rightarrow $ 2 Born scattering (single graviton exchange) of  massive scalars in ${\mathbb R}^{1,D-2} \times S^1$. We then compactify on ${\mathbb R}^{1,D-2} \times S^1$  by quantizing the momentum along the circle as usual. We show how the resulting amplitude  can be interpreted in ${\mathbb R}^{1,D-2}$ as the scattering of 2 charged scalar fields with Kaluza-Klein  masses $m_1 $ and $m_2$ via single graviton, photon and dilaton exchange. We also perform  consistency checks on the amplitude to show how we may recover the expected results in the $R \rightarrow \infty $ and $R \rightarrow 0 $ limits where $R$ is the radius of the compactified circle. In section \ref{doubleexchange} we calculate the one-loop contribution to the  2 $\rightarrow $ 2  scattering in ${\mathbb R}^{1,D-2} \times S^1$ using standard methods of introducing Schwinger parameters for the internal propagators. We explicitly consider this scattering amplitude in the Regge high energy limit with $s \gg |t| $ whilst keeping $t$ fixed, which is interpreted as the limit $s' \gg |t'| $ with Kaluza-Klein masses held fixed, where $s'$, $t'$ are the Mandelstam variables of the theory on ${\mathbb R}^{1,D-2} $. In going to impact parameter space we find that corresponding amplitude is related to the square of the tree-level expression in precisely the way required for the eikonal phase to exponentiate. There is however a technical limitation in our proof which requires us to restrict to the case $D=5$. In section \ref{kinlimits} we look at various limits of the leading order eikonal expression  alluded to above. We also relate the subleading eikonal in the Kaluza-Klein theory in a particular limit to the subleading eikonal from scattering massless states from a stack of $D$p-branes. In section \ref{d0d6dual} we draw some further parallels and insights between massless scalar scattering off heavy Kaluza-Klein modes in our model and that of dilaton scattering off stacks of $D0$-branes in $D=4$ or $D6$-branes in $D=10$ in IIA string theory. In section \ref{disc} we discuss our results.
\clearpage

\section{Kaluza-Klein Theory} \label{kktheory}

In this section we briefly consider some generalities relating to Kaluza-Klein theory \cite{Kaluza:1921tu,Overduin:1998pn} which we will use throughout the paper. For definiteness lets first consider the case $D=5$ and thus ${\mathbb R}^{1,4}$ compactified to ${\mathbb R}^{1,3} \times S^1  $. We first need to identify the correct charges of the massive Kaluza-Klein scalars. We start with the $D=5$ gravity action coupled to a real massless scalar field,
\be
S_5 = \int d^5x  \sqrt{-\hat{g} } \, \left( \frac{1}{2 \kappa_5^2} {\hat R}   - \frac{1}{2} \partial_M \Phi \partial_N \Phi {\hat g}^{MN} \right) \;,
\label{eq:5DKKaction}
\ee
with indices $M,N=0 \ldots 4 $ and $\kappa_5^2 = 8 \pi G_5 $, ${\hat{g} }  = \det ({\hat g}_{MN} ) $. Note that the hat denotes $D=5$ quantities. Taking the standard form for the Kaluza-Klein $D=5$ line element (with coordinate $x^4 $  parametrizing the circle of radius $R$),
 \be
 d {\hat S}^2 = \phi^{-1/3}g_{\mu \nu} dx^\mu dx^\nu + \phi^{2/3}(dx^5 + \lambda A_\mu dx^\mu )^2  \;,
 \label{eq:5DKKmetric}
 \ee
where $g_{\mu \nu}, A_\mu $ and $\phi$ denote the $D=4$ graviton, $U(1)$ gauge field and dilaton respectively. Substituting the metric given by \eqref{eq:5DKKmetric} into \eqref{eq:5DKKaction}, including both the $D=4$ massless modes arising from the $D=5$ metric as well as the infinite tower of massive Kaluza-Klein states arising from $\Phi$ we find that the $D=4$ action \cite{Blau} is given by,
\begin{eqnarray}
S_4 &=&  \int d^4x  \sqrt{- g } \, \biggl(  \frac{1}{2 \kappa_4^2} { R}   -\frac{1}{4}\phi  F_{\mu \nu}F^{\mu \nu} -  \frac{1}{96 \pi G_4} \frac{1}{\phi^2} \partial_\mu \phi \partial^\mu \phi   \nonumber \\
&& - \frac{1}{2} \sum_{n \in {\mathbb Z }} \bigl( ( D_\mu \Phi_n )( D^\mu \Phi^{*}_n  ) + m_n^2 \phi^2  \Phi^{*}_n  \Phi_n \bigr) \biggr) \;,
\end{eqnarray}
where  $k_4^ 2 = 8 \pi G_4 $, we note that the $D=5$ and $D=4$ gravitational constants are related by $G_5 = 2 \pi R G_4 $ and the parameter $\lambda $ appearing in the $D=5$ metric ansatz \eqref{eq:5DKKmetric} is fixed to be $\lambda = \sqrt{16\pi G_4 }$ in order to obtain the  canonical form of the Maxwell term in the $S_4$ action.
 
It is conventional to redefine the dilaton $\phi = e^{\sigma \sqrt{48\pi G_4}}$  which leads to a canonical kinetic term for $\sigma$. With this substitution we find,
\begin{eqnarray}
\label{KK4D}
S_4 &=&  \int d^4x  \sqrt{- g } \, \biggl(  \frac{1}{2 \kappa_4^2} { R}   -\frac{1}{4} e^{\sigma \sqrt{48\pi G_4}}  F_{\mu \nu}F^{\mu \nu}  -   \frac{1}{2}  \partial_\mu \sigma \partial^\mu \sigma  \nonumber \\
&& - \frac{1}{2} \sum_{n \in {\mathbb Z }} \bigl( ( D_\mu \Phi_n )( D^\mu \Phi^{*}_n  ) +  e^{-\sigma \sqrt{48\pi G_4}} m_n^2  \Phi^{*}_n  \Phi_n \bigr) \biggr) \;.
\end{eqnarray}
Expanding the terms in the action about $\sigma = 0 $ we see that the $\Phi_n$ fields are massive, charged complex scalars, with $ D_\mu \Phi_n  = ( \partial_\mu \Phi_n -iQ_n A_\mu \Phi_n ) $ and charges given by $Q_n  =n  \sqrt{16 \pi G_4}/R $.

\section{Scattering in the Born approximation} \label{bornapprox}

\subsection{Scattering on ${\mathbb R}^{1,D-1} $} \label{TL}

\begin{figure}[h]
  \centering
  \begin{tikzpicture}[scale=1.5]
	    \begin{feynman}
			\vertex (a) at (-2,-2) {};
			\vertex (b) at ( 2,-2) {};
			\vertex (c) at (-2, 0) {};
			\vertex (d) at ( 2, 0) {};
			\vertex[circle,inner sep=0pt,minimum size=0pt] (e) at (0, 0) {};
			\vertex[circle,inner sep=0pt,minimum size=0pt] (g) at (0, -2) {};
			\diagram* {
			(c) -- [fermion,edge label=$k_1$] (e) -- [anti fermion,edge label=$k_3$] (d),
			(g) -- [boson] (e),
			(a) -- [fermion,edge label=$k_2$] (g) -- [anti fermion,edge label=$k_4$] (b),
			};
	    \end{feynman}
  \end{tikzpicture}
  \caption{Feynman diagram representation for tree-level scalar scattering with graviton exchange. The solid lines represent scalar states and the wavy lines represent gravitons.}
  \label{fig:treegrav}
\end{figure}
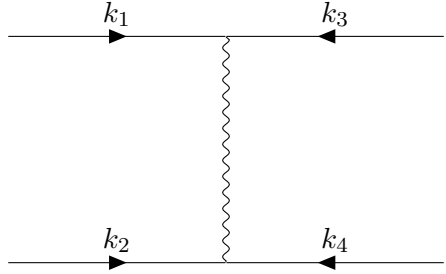

In this subsection we first review the tree-level scattering of two massive scalar fields  with masses $M_1, M_2$, incoming momenta $p_1, p_2$ and outgoing momenta $p_3, p_4$ via single graviton exchange as illustrated in figure \ref{fig:treegrav}. The resulting momentum space amplitude is,

\begin{equation}
  \label{eq:singlegraviton}
   i{\cal A}_0 (D)= 2i \kappa_D^2 \frac{1}{ (p_1+p_3)^2} \biggl( \frac{1}{2} (s-M_1^2-M_2^2)^2 - \frac{2}{D-2} M_1^2 M_2^2 +\frac{1}{2}(s -M_1^2 -M_2^2)t \biggr) \;,
\end{equation}
where our Mandelstam variables conventions are  $s = - (p_1+p_2)^2 , t = -(p_1+p_3)^2, u = -(p_1+p_4)^2 $,  working in $D$ non-compact spacetime dimensions ${\mathbb R}^{1,D-1} $ with mostly positive signature. Note here that $\kappa_D = \sqrt{8 \pi G_D} $ with $G_D$  the gravitational constant in $D$ non-compact space time dimensions.

In the large $s$, $M_1$, $M_2$ fixed $t$ high energy limit and neglecting the linear term in $t$ which does not contribute to the eikonal phase at large impact parameter $b$, we find

\begin{equation}
  \label{eq:}
   i\mathcal{A}_{0} =  \frac{2 i \kappa_D^2 \gamma(s)}{q^2} \;,
\end{equation}
where we have defined $\gamma(s)=2(k_1 \cdot k_2)^2 - \frac{2}{D-2} k_1^2 k_2^2 = \frac{1}{2} (s-M_1^2-M_2^2)^2 - \frac{2}{D-2} M_1^2 M_2^2$. If we now consider the above amplitude in impact parameter space, we define the momentum transfer as $q_\perp \equiv  (p_1+p_3)$ lying in the impact plane with $b$, the impact parameter, defined as the corresponding conjugate variable. This allows us to define the eikonal phase $\chi$ 
\begin{equation}
  \label{eq:defneikonal}
   \chi (D) =  { \tilde{\cal A}}_0 = \frac{1}{4Ep} \int \frac{d^{D-2}q_\perp}{(2\pi )^{D-2}}  e^{i q_\perp \cdot b }   {\cal A}_0 \;,
\end{equation}
where $E=E_1+E_2$. In the Regge high energy limit one may approximate $t = - q_\perp^2 $ thus allowing the integrals over the $D-2$ transverse momenta to be carried out. The final result is 
\begin{equation}
i \chi (D) = i \tilde{\mathcal{A}}_{0}(D) = \frac{i \kappa_D^2 \gamma(s)}{2 E p} \frac{1}{4 \pi^{\frac{D-2}{2}}} \Gamma \left( \frac{D}{2}-2 \right) \frac{1}{b^{D-4}} \;, \label{TLNC}
\end{equation}
where $(E_1+E_2)p = \sqrt{(k_1 \cdot k_2)^2 - k_1^2 k_2^2}$ which we can see by defining $k_1=(E_1,\ldots,p)$ and $k_2=(E_2,\ldots,-p)$.

\subsection{Scattering on ${\mathbb R}^{1,D-2} \times S^1$} \label{treeKKscat}

In this subsection we want to consider the case of large $s$ fixed $t$ eikonal scattering but where one of the spatial transverse directions has been compactified on a circle of radius $R$. The calculation is similar except that we have to replace an integration over continuous momentum with a sum over discrete quantized momentum along the $S^1$ when Fourier transforming the amplitude into impact parameter space. Let us define $q_\perp = ({\bf q}_\perp', q_s)$ with $q_s = n/R $, $n \in \mathbb{Z}$ being the integer valued momentum number. Correspondingly, we partition the impact parameter $b$ previously defined over $D-2$ transverse directions into its components along the $D-3$ transverse directions and the circle $S^1 $ so we define $b = ( {\bf b}', b_s ) $. It will also be useful to define the new Mandelstam variable $s'$  pertaining to $D-1$ dimensional spacetime,  so $s' = -(p_1'+p_2')^2 $ where $'$ denotes momenta restricted to $ {\mathbb R}^{1,D-2}$. We thus have the relation $ s = s' - \frac{(n_1 +n_2)^2}{R^2} $.

We note that in order to get the corresponding impact parameter space amplitude in ${\mathbb R}^{1,D-2} \times S^1$ from \eqref{eq:defneikonal} we need the following,

\begin{equation}
  \label{eq:changeconttoKK}
     \int d^{D-2}q_\perp \rightarrow \int d^{D-3}q'_\perp \frac{1}{R}  \sum_{n  \in {\mathbb Z } } \;.
\end{equation}
Using the fact that  $q_\perp$ and $q_\perp'$ are related via $ q_\perp^2 = (q_\perp')^2 + (n/R)^2 $, the impact parameter space amplitude for scattering on 
${\mathbb R}^{1,D-2} \times S^1  $ can therefore be written as, 
\ba
  \label{eq:compeikonal}
i { \tilde{\cal A}}_0 (D-1,R)  &= &i\frac{\kappa^2 (s' - m_1^2 - m_2^2 - 2 m_1 m_2)}{2} \int  \frac{d^{D-3}q_\perp'}{(2\pi )^{D-2}} e^{i q_\perp' \cdot b' }  \frac{1}{R}  \sum_{n  \in {\mathbb Z } } \frac {e^{ib_s n/R}}{(q_\perp')^2 + (n/R)^2 }\cr
 \cr
&\equiv & i\chi(D-1,R) \;,
\ea
where we have defined $m_1^2 = (n_1/R)^2 $ and $m_2^2 = (n_2/R)^2 $ which we recognise as the Kaluza-Klein masses associated with compactification on $S^1$ which are different from the capital $M$ masses used in the previous section. Note that we have taken the 5-dimensional masses to be 0 as suggested by \eqref{eq:5DKKaction}. We also notice that $n=n_1+n_3$. The integration over the $q_\perp' $ momenta can be carried out explicitly giving,
\begin{equation}
  \label{TLC}
 i    { \tilde{\cal A}}_0 (D-1, R)  = i\frac{\kappa^2 }{2} \frac{s' - m_1^2 -m_2^2 - 2 m_1 m_2  }{(2\pi)^{\frac{D-1}{2}} }\frac{1}{ {(b')}^{\frac{D-5}{2} }}  \frac{1}{R}  \sum_{n  \in {\mathbb Z } } e^{ib_s n/R}  {\biggl( \frac{|n|}{R} \biggr)^{\frac{D-5}{2} } K_{\frac{D-5}{2} } \biggl( \frac{|n|b'}{R} \biggr) \;,
}\end{equation}
where $K_n(x) $ is a Bessel K function. In the ultra-relativistic limit $ s' \gg m_1^2, m_2^2  $ we find,
\be
\label{RTLC}
i { \tilde{\cal A}}_0 (D-1, R)  = i\frac{\kappa^2 }{2} \frac{s'}{(2\pi)^{\frac{D-1}{2}} }\frac{1}{ {(b')}^{\frac{D-5}{2} }}  \frac{1}{R}  \sum_{n  \in {\mathbb Z } } e^{ib_s n/R}  {\Bigl( \frac{|n|}{R} \Bigr)^{\frac{D-5}{2} } K_{\frac{D-5}{2} } \Bigl( \frac{|n|b'}{R} \Bigr) } \;.
\ee
We can check the consistency of \eqref{RTLC} by considering the two limits $R \rightarrow 0 $  and $R \rightarrow  \infty $ in which we expect to recover the previous expression \eqref{TLNC} with $D-1$ and $D$  spacetime dimensions respectively.

First we consider the limit $R \rightarrow 0 $. In this limit we expect only the  zero momentum $n=0$ Kaluza-Klein mode to contribute to the sum over t-channel states. This can formally be seen using the 
known expansion of the Bessel K functions $ K_\nu (z)  \sim \sqrt{\pi /{2z} } e^{-z} +.... $ as $z \rightarrow \infty $ which applies when $n\neq 0 $.  For the  $n=0 $ contribution we have to consider $ K_\nu (z) $ for $z  \rightarrow  0 $. The behaviour in this limit is,
\begin{equation}
\label{Kz0}
K_\nu (z)  = \frac{1}{2} \Bigl[ \Gamma(\nu) \Bigl( \frac{z}{2} \Bigr)^{-\nu } (1 + \frac{z^2}{4(1+\nu ) } + ....)  +  \Gamma(-\nu) \Bigl( \frac{z}{2} \Bigr)^{\nu } (1 + \frac{z^2}{4(1+\nu ) } + ....)  \Bigr] \;.
\end{equation}
Hence we find,
\begin{equation}
\lim_{n \rightarrow 0} \, \left[ \left( \frac{|n|}{R} \right)^{\frac{D-5}{2}} K_{\frac{D-5}{2}} \left( \frac{|n|b'}{R} \right) \right] =  \left( \frac{1}{2} \right)^{\frac{7-D}{2}} \Gamma \left(\frac{D-5}{2} \right) \frac{1}{(b')^{\frac{D-5}{2}}} \;.
\end{equation}   
Using this, the final result  for the amplitude  $ \tilde{\cal A}_0 (R)  $ as $R \rightarrow 0 $ is
\begin{equation}
 \label{AR0}
 \lim_{R \rightarrow 0} \, [ i { \tilde{\cal A}}_0 (D-1,R) ] = i\frac{s'}{2} G_{D-1} \frac{ \Gamma \bigl( \frac{D-5}{2}  \bigr) }{\pi^{\frac{D-5}{2}} b'^{D-5} } \;,
 \end{equation} 
where $G_{D-1} $ is the  gravitational constant in the $D-1$ non-compact spacetime dimensions and is related to  the one in $D$ dimensions via the familiar Kaluza-Klein  relation $G_D$ via $ G_D = 2\pi R G_{D-1} $. It is clear that the right hand side of the expression \eqref{AR0}  is the same as $\tilde{\cal A}_0 (D-1)$ and the corresponding eikonal phase $ \chi(D-1,R \rightarrow 0) = \chi(D-1)$ where these quantities have been discussed in section \ref{TL}. So we recover the expected result,  namely  the usual expression for high energy tree-level scattering but with $D-1$ non-compact spacetime dimensions.
 
We now consider the opposite limit of $\chi(D-1, R) $ with $R \rightarrow \infty $ where we should recover the expression $\chi(D)$ of \eqref{TLNC}. Noting that in this limit the discrete sum over momenta becomes a continuous integral $ \frac{1}{2 \pi R} \sum_{n} \rightarrow \frac{1}{2\pi} {\int}_{-\infty}^{\infty} dq $ the amplitude becomes
 \begin{equation}
  \label{ }
  i{ \tilde{\cal A}}_0 (D-1, R \rightarrow \infty )  =   i\frac{4 \pi G_D s  }{ (2 \pi)^{\frac{D-1}{2}} (b')^{D-4}  } I(b_s/b' ) \;,
 \end{equation}
where we use the fact that $s' \rightarrow s $ as $R \rightarrow \infty $ and the integral $I(b_s/b' ) $  is given by 
\begin{equation}
I(b_s/b' ) = {\int}_{-\infty}^{\infty} d{\tilde q } \, e^{i {\tilde q }(b_s/b' ) }   \bigl( |{\tilde q}| \bigr)^{\frac{D-5}{2}}   K_{\frac{D-5}{2}} \bigl( |{\tilde q}| \bigr) \;,
 \end{equation}
 with ${\tilde q} = q \,b' $. The integral $ I(b_s/b' )  $ can be computed using the cosine integral transform formula for the  function $  {\tilde q}^\nu K_\nu({\tilde q}  ) $,
\begin{equation}
\int _0^\infty dx \, {\rm cos}(xy) x^{\pm \mu } K_\mu (a x) = \sqrt{\pi/4} \, \bigl(2a\bigr)^{\pm \mu} \Gamma\left( \pm\mu +\frac{1}{2} \right) ( y^2 +a^2 )^{\mp \mu -\frac{1}{2}  } \;,
 \end{equation}
with $ \mu > -\frac{1}{2} $ for the upper sign and $ \mu < \frac{1}{2}  $ for the lower sign, with  ${\rm Re}(a) > 0 $. In our case, $\mu = (D-5)/2$, $a=1 $ and $y = b_s/b'$. Taking $D>5 $ we find
\begin{equation}
i { \tilde{\cal A}}_0 (D-1, R \rightarrow \infty )  = i \frac{4\pi G_D \, s }{ (2 \pi)^{\frac{D-1}{2}} (b')^{D-4}  } \frac{  {(2\pi )}^{\frac{D-5}{2} } \Gamma\bigl( \frac{D-4}{2} \bigr)} {\bigl( b_s^2/b'^2 +1 \bigr)^{\frac{D-4}{2}} } \;.
\end{equation}
Using the relation  between the impact parameters $b,  b' $ and $b_s $,   $ b^2 = b'^2 +b_s^2 $   we can see that 
\begin{equation}
i { \tilde{\cal A}}_0 (D-1, R \rightarrow \infty ) = { i \tilde{\cal A}}_0 (D) \;,
\end{equation}
which is the same as \eqref{TLNC} in the ultra-relativistic limit as expected.

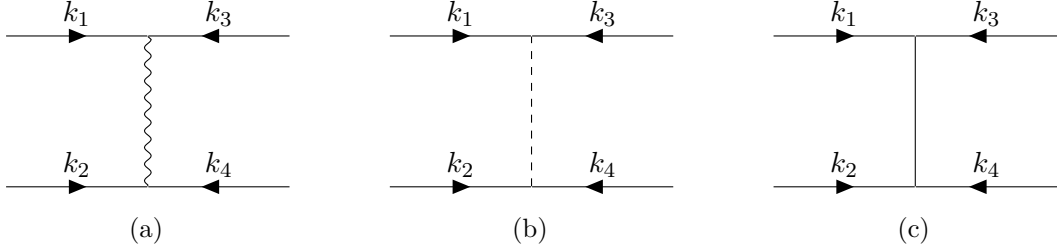
\begin{figure}[h]
  \begin{subfigure}[t]{0.3\textwidth}
    \centering
    \begin{tikzpicture}
	    \begin{feynman}
			\vertex (a) at (-2,-2) {};
			\vertex (b) at ( 2,-2) {};
			\vertex (c) at (-2, 0) {};
			\vertex (d) at ( 2, 0) {};
			\vertex[circle,inner sep=0pt,minimum size=0pt] (e) at (0, 0) {};
			\vertex[circle,inner sep=0pt,minimum size=0pt] (g) at (0, -2) {};
			\diagram* {
			(c) -- [fermion,edge label=$k_1$] (e) -- [anti fermion,edge label=$k_3$] (d),
			(g) -- [boson] (e),
			(a) -- [fermion,edge label=$k_2$] (g) -- [anti fermion,edge label=$k_4$] (b),
			};
	    \end{feynman}
    \end{tikzpicture}
    \caption{}
    \label{fig:3a}
  \end{subfigure}
  \quad
  \begin{subfigure}[t]{0.3\textwidth}
    \centering
    \begin{tikzpicture}
	    \begin{feynman}
			\vertex (a) at (-2,-2) {};
			\vertex (b) at ( 2,-2) {};
			\vertex (c) at (-2, 0) {};
			\vertex (d) at ( 2, 0) {};
			\vertex[circle,inner sep=0pt,minimum size=0pt] (e) at (0, 0) {};
			\vertex[circle,inner sep=0pt,minimum size=0pt] (g) at (0, -2) {};
			\diagram* {
			(c) -- [fermion,edge label=$k_1$] (e) -- [anti fermion,edge label=$k_3$] (d),
			(g) -- [scalar] (e),
			(a) -- [fermion,edge label=$k_2$] (g) -- [anti fermion,edge label=$k_4$] (b),
			};
	    \end{feynman}
    \end{tikzpicture}
    \caption{}
    \label{fig:3b}
  \end{subfigure}
  \quad
  \begin{subfigure}[t]{0.3\textwidth}
    \centering
    \begin{tikzpicture}
	    \begin{feynman}
			\vertex (a) at (-2,-2) {};
			\vertex (b) at ( 2,-2) {};
			\vertex (c) at (-2, 0) {};
			\vertex (d) at ( 2, 0) {};
			\vertex[circle,inner sep=0pt,minimum size=0pt] (e) at (0, 0) {};
			\vertex[circle,inner sep=0pt,minimum size=0pt] (g) at (0, -2) {};
			\diagram* {
			(c) -- [fermion,edge label=$k_1$] (e) -- [anti fermion,edge label=$k_3$] (d),
			(g) -- (e),
			(a) -- [fermion,edge label=$k_2$] (g) -- [anti fermion,edge label=$k_4$] (b),
			};
	    \end{feynman}
    \end{tikzpicture}
    \caption{}
    \label{fig:3c}
  \end{subfigure}
  \caption{The various constituent diagrams found in tree-level scalar scattering with graviton exchange on ${\mathbb R}^{1,D-2} \times S^1$ when decomposed into seperate dilaton, gauge field and dimensionally reduced graviton contributions. The solid lines represent scalar states, the dashed line represent gauge fields and the wavy lines represent gravitons. Note that the internal solid lines represent massless dilaton states.}
  \label{fig:treegravcompact}
\end{figure}

In the above we have  focussed on the ultra-relativistic limit $ s' \gg m_1^2, m_2^2  $ . We now wish to consider relaxing this  condition  so that later on we can consider the comparison of scattering amplitudes and eikonal phases in the Kaluza-Klein theory with those previously studied in \cite{Collado:2018isu} which included massless dilatons elastically scattering off a large stack of $Dp$-branes. If we return to  \eqref{eq:compeikonal} and consider the momentum space amplitude from which it derives, we have,

\be 
 i {{\cal  A}}_0 (D-1, R ) = -i 2\pi R\kappa_5^2\frac{ (s' - m_1^2 -m_2^2 - 2 m_1 m_2 )^2 }{t' - (n_1 +n_3)^2/R^2 } \;.
\ee
This corresponds to a single $D=5$ graviton exchange between the massive states $\Phi_n $ in the Kaluza-Klein tower arising from compactification on $ {\mathbb R}^{1,D-2} \times S^1$ where we recall that $m_1^2 = n_1^2/R^2$, $m_2^2 = n_2^2/R^2 $. This scattering is  inelastic if $n_1 \neq  -n_3  $  in which case massive $D=4$ states are exchanged. We wish to focus here on elastic scattering so we will chose kinematics such that $n_1 = -n_3 $ (and hence $ n_2 = -n_4 $) but both $n_1 $ and $n_2 $ non-zero in general. Then $ {{\cal  A}}_0 (D-1, R ) $ may be written as,
\be 
 i {{\cal  A}}_0 (D-1, R ) = -2i \frac{\kappa_4^2 }{t'} \left(\frac{1}{2} (s' - m_1^2 -m_2^2 )^2 - 2 m_1m_2 (s' - m_1^2 -m_2^2)  + 2 m_1^2 m_2^2  \right) \;.
\ee
The first term in the round brackets is almost of the form of the diagram involving  a single massless graviton exchange in $D=4$ between two massive scalars, as  given in \eqref{eq:singlegraviton} with choice $D=4$. Therefore it is useful to rewrite ${{\cal  A}}_0 (D-1, R )$ in the form,
\be
\label{KKdecomp}
i {{\cal  A}}_0 (D-1, R ) = -2i \frac{\kappa_4^2 }{t'} \left(\frac{1}{2} [ (s' - m_1^2 -m_2^2 )^2 - m_1^2m_2^2 ] - [ 2 m_1m_2 (s' - m_1^2 -m_2^2)]   + 3 m_1^2 m_2^2  \right) \;.
\ee
Now the first term in the square brackets is precisely the  contribution from a single $D=4$ graviton. Based on the fact  that a massless $D=5$ graviton gives rise to a massless graviton, photon and dilaton in $D=4$ (along with their massive Kaluza-Klein  excitations) we would expect that the term in the second square brackets to correspond to massless photon exchange and finally that the last term to correspond to massless dilaton exchange. The several Feynman diagrams are shown in figure \ref{fig:treegravcompact}.

Let us check that this is the case. The diagram for single photon exchange via the t-channel is easily computed in $D=4$ scalar electrodynamics. Given two charged massive scalars of masses $m_1, m_2$ and electric charges $q_1, q_2$, the resulting momentum space amplitude, $\mathcal{A}_{\text{phot}}$, is found to be, 
\be
i \mathcal{A}_{\text{phot}} = 2i q_1 q_2 \frac{( s' -m_1^2 -m_2^2)}{t'} \;.
\ee
Now in the Kaluza-Klein case we previously read off the charges from the action and so we can write $q_1 = \sqrt{2} \kappa_4 n_1/R $ and  $q_2 = \sqrt{2}\kappa_4 n_2/R $  giving 
\be
i \mathcal{A}_{\text{phot}} = 4i \kappa_4^2 m_1 m_2 \frac{( s' -m_1^2 -m_2^2)}{t'} \;,
\ee
which is precisely the contribution of the second square brackets in \eqref{KKdecomp}.

Finally we need to show that the final term in \eqref{KKdecomp} corresponds to a single dilaton exchange in $D=4$. This is straightforward as it only depends on the coupling between the massive Kaluza-Klein scalars $\Phi_n$ and the dilaton $\sigma$ which from the action \eqref{KK4D} is found to be $ -i \sqrt{6} \kappa_4 m_n^2 $. The contribution, $\mathcal{A}_{\text{dil}}$ is thus,
\be 
i \mathcal{A}_{\text{dil}} = -6 i \kappa_4^2 \frac{m_1^2 m_2^2}{t'} \;,
\ee
which is exactly the last term in \eqref{KKdecomp}.

\section{Double Exchange Scattering} \label{doubleexchange}

\subsection{Scattering on ${\mathbb R}^{1,D-1}$} 

\begin{figure}[h!]
  \centering
  \begin{tikzpicture}[scale=1.5]
	    \begin{feynman}
			\vertex (a) at (-2.5,-2) {};
			\vertex (b) at ( 2.5,-2) {};
			\vertex (c) at (-2.5, 0) {};
			\vertex (d) at ( 2.5, 0) {};
			\vertex[circle,inner sep=0pt,minimum size=0pt] (e) at (-1, 0) {};
			\vertex[circle,inner sep=0pt,minimum size=0pt] (f) at (1, 0) {};
			\vertex[circle,inner sep=0pt,minimum size=0pt] (g) at (-1, -2) {};
			\vertex[circle,inner sep=0pt,minimum size=0pt] (h) at (1, -2) {}; 
			\diagram* {
			(c) -- [fermion,edge label=$k_1$] (e) -- [fermion] (f) -- [anti fermion,edge label=$k_3$] (d),
			(g) -- [boson] (e),
			(h) -- [boson] (f),
			(a) -- [fermion,edge label=$k_2$] (g) -- [fermion] (h) -- [anti fermion,edge label=$k_4$] (b),
			};
	    \end{feynman}
  \end{tikzpicture}
  \caption{Feynman diagram representation for one-loop scalar scattering with two graviton exchanges. The solid lines represent scalar states and the wavy lines represent gravitons.} \label{fig:loopgrav}
\end{figure}
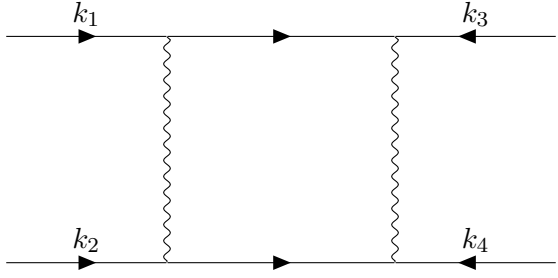

In this subsection we will briefly calculate the one-loop double exchange scattering amplitude, shown in figure \ref{fig:loopgrav}, in normal flat space. We will remain general here and keep the masses of the two scalars non-zero and large. In order to more readily calculate the diagram in the high energy limit we will use the following approximation to the scalar-scalar-graviton vertex \cite{Giddings:2011xs, Kabat:1992tb},
\begin{equation}
-i \kappa \left (p_{\mu}p_{\nu}' + p_{\nu}p_{\mu}' - (p \cdot p' - m^2) \eta_{\mu \nu} \right ) \approx -2i\kappa_D p_{\mu}p_{\nu} \mbox{ ,}
\end{equation}
where we have taken the incoming momenta $p$ to be roughly equal to the outgoing momenta $-p'$ and have used the mass-shell condition. With this approximation to the vertex we can write the one-loop amplitude as,
\begin{equation}
i \mathcal{A}_{1}  =  4 \kappa_D^4 \gamma^2(s) \int \frac{{\textrm{d}}^D k}{(2\pi)^D} \frac{1}{k^2} \frac{1}{(q+k)^2} \frac{1}{(k_1 + k)^2 + M_1^2} \frac{1}{(k_2 - k)^2 + M_2^2} \mbox{ .} \label{ssg-loop-2}
\end{equation}

\noindent
The integral can be solved and in the high energy limit, we find,
\begin{equation}
\mathcal{I}_4 = \frac{i \pi^{\frac{D}{2}}}{(2\pi)^D} \frac{\pi}{4} \frac{1}{\sqrt{M_1^2 M_2^2 - (k_1 \cdot k_2)^2 }} \frac{\Gamma^2(\frac{D}{2}-2) \Gamma(3-\frac{D}{2})}{\Gamma(D-4)} (q^2)^{\frac{D}{2}-3} + \ldots
\end{equation}
Putting this together we find that,
\begin{equation}
i \mathcal{A}_{1} \approx  \frac{i \pi^{\frac{D}{2}}}{(2\pi)^D} \frac{\pi}{4} \frac{4 \kappa_D^4 \gamma^2(s)}{\sqrt{M_1^2 M_2^2 - (k_1 \cdot k_2)^2 }} \frac{\Gamma^2(\frac{D}{2}-2) \Gamma(3-\frac{D}{2})}{\Gamma(D-4)} (q^2)^{\frac{D}{2}-3} \;,
\end{equation}
where we have neglected contributions that are subleading. Moving to impact parameter space as before we find that,
\begin{equation}
i \tilde{\mathcal{A}}_{1} = - \frac{\kappa_D^4 \gamma^2(s)}{(Ep)^2} \frac{1}{2^7 \pi^{D-2}} \Gamma^2 \left(\frac{D}{2}-2 \right) \frac{1}{b^{2D-8}} \;.
\end{equation}
Comparing with \eqref{TLNC} we easily see that $i \tilde{\mathcal{A}}_{1}=\frac{1}{2} (i \tilde{\mathcal{A}}_{0})^2$ as expected. Note that we have included both the contributions from the regular and crossed double exchange diagrams.

\subsection{Scattering on ${\mathbb R}^{1,D-2} \times S^1$} 
 
We want to compute the leading one-loop contribution to the scattering process considered in subsection \ref{treeKKscat}. In doing so we will extend the results in the previous subsection to scattering on ${\mathbb R}^{1,D-2} \times S^1$. We will compute the one-loop amplitude in the ultra-relativistic limit but the results should easily extend to case where both Kaluza-Klein masses are kept large.

The method we use is the standard one of introducing Schwinger parameters $t_i$ for $i = 1\ldots4$ corresponding to the 4 internal propagators of the diagram. Starting from \eqref{ssg-loop-2}, using \eqref{eq:changeconttoKK} and making a change of variables from $t_i $ to $T, t_2,t_3, t_4$ where $T = \sum_i t_i $ we can express the one-loop amplitude in the ultra-relativistic limit as,
 \begin{equation}
i { {\cal A}}_1 (D-1, R  )  =  4i \kappa^4 (s'^2)^2 \frac{1 }{(2\pi)^{D-1}} \int _0^\infty dT\, {\displaystyle  \prod_{i=2}^4 }\int _0^\infty dt_i  \int  \frac{ d^{D-1} q'}{(2\pi)^{D-1}} \frac{1}{2\pi R} \,  \sum_{n  \in {\mathbb Z } } e^{-T\bigl( q + A/2T \bigr)^2 } \;,
 \end{equation}
 where   we use the notation $  \bigl( q + A/2T \bigr)^2 = \bigl( q' + A'/2T \bigr)^2 + \bigl( n/R+ A_s/2T \bigr)^2   $  with    $q = (q', n/R) $ being  the loop momentum on  ${\mathbb R}^{1,D-2} \times S^1  $
 and 
 $A = (A', A_s) $ with 
 \begin{eqnarray}
 A' &=& (- 2p_3't_2 - 2p_1't_2 -2 p_1't_3 +2 p_2' t_4) \cr 
 \cr
A_s &=& (- 2n_3 t_2 - 2n_1 t_2 -2 n_1 t_3 +2 n_2 t_4)/R \;.
  \end{eqnarray}
 The integral over the $q'$ momenta is just a Gaussian integral after shifting variables $ q' \rightarrow q' - A'/2T $  so that,
  \begin{equation}
i { {\cal A}}_1 (D-1, R  )  =  \frac{4i \kappa^4 (s'^2)^2 }{2 \pi R} \frac{\pi ^{\frac{D-1}{2} }}{(2\pi)^{D-1}} \int _0^\infty dT  \,  T^{-{\frac{D-1}{2}}}{\displaystyle  \prod_{i=2}^4 }\int _0^\infty dt_i   \,  \sum_{n  \in {\mathbb Z } } e^{-T\bigl( n/R + A_s/2T \bigr)^2} e^{\frac{f}{T} +t_2 t}  \;,
 \end{equation}
where we have defined, $f = t(-t_2^2 -t_2 t_3 -t_2t_4 ) + t_3t_4s$. Changing to new variables $ \alpha_i \equiv t_i/T $,
\begin{equation}
\label{A1loop}
i{ {\cal A}}_1 (D-1, R  )  = \frac{4i \kappa^4 (s'^2)^2 }{2 \pi R} \frac{\pi ^{\frac{D-1}{2} }}{(2\pi)^{D-1}} \int _0^\infty dT  \,  T^{3-{\frac{D-1}{2}}}{\displaystyle  \prod_{i=2}^4 }\int _0^\infty d\alpha_i   \,  \sum_{n  \in {\mathbb Z } } e^{-T\bigl( n/R + {\tilde A}_s  \bigr)^2} e^{T {\tilde f} } \;,
\end{equation}
with  $ {\tilde A}_s = (- n_3 \alpha_2 - n_1 \alpha_2 - n_1 \alpha_3 + n_2 \alpha_4)/R  $  and $ {\tilde f} =  t(-\alpha_2^2  - \alpha_2 \alpha_3 -\alpha_2 \alpha_4 )  +\alpha_3 \alpha_4 s + \alpha_2 t$. If we combine the $n$ independent terms that multiply $T$ in the exponential term in \eqref{A1loop}, we find after some algebra that this can be expressed as, 
  \begin{eqnarray}
 &-&\alpha_2^2 t'  -\alpha_2 \alpha_3 \bigl( t' - (n_3+n_1)^2/R^2 +n_1(n_1+n_3)/R^2 \bigr) - \alpha_2 \alpha_4 \bigl(t'  -(n_3+n_1)^2/R^2 \cr
 \cr
 &-&n_2(n_3+n_1)/R^2 \bigr)  -\alpha_3^2  n_1^2/R^2  - \alpha_4^2 n_2^2/R^2  - \alpha_3 \alpha_4 \bigl(  n_1^2/R^2 + n_2^2/R^2  \bigr) \cr
 \cr
 &+&  \alpha_3 \alpha_4 s' + \alpha_4 t'- \alpha_2 (n_3+n_1)^2 /R^2 \;.
  \end{eqnarray}
The appearance of the Mandelstam variables $s', t'$ and the mass terms $m_1^2 = n_1^2/R^2 ,  m_2^2 = n_1^2/R^2 $  in the coefficients of $\alpha_3^2 $ and  $\alpha_4^2 $ mean the expression above is what we would expect from scattering of massive Kaluza-Klein modes in ${\mathbb R}^{1,D-2}$.
 
The original high energy limit $s \rightarrow \infty $ with $t$ kept fixed, can be understood in the compactified case as the limit $s' \rightarrow \infty$ with $t'$ fixed and the Kaluza-Klein momenta $n_i/R$, $i = 1\ldots4 $ fixed. Now in the $s' \rightarrow \infty$ limit, the only non-vanishing contributions in the $\alpha_3, \alpha_4 $ integrals come from the regions $\alpha_3, \alpha_4  \rightarrow 0 $ so as to keep the term in the exponential from diverging. These integrals may then be solved after a wick rotation in  ${\mathbb R}^{1,D-2}$ which yield Gaussian integrals resulting in a factor $  \frac{\pi}{2 T \sqrt{-{E'_e}^4} }$ where $E'_e $ is the corresponding energy in Euclidean space. Combining these results we find that the amplitude then reduces to, 
 \begin{eqnarray}
i{  { \cal A}}_1 (D-1, R  )  &= &\frac{4i \kappa^4 (s'^2)^2 }{2 \pi R} \frac{\pi ^{\frac{D+1}{2} }}{(2\pi)^{D-1}}  \frac{1}{2 T \sqrt{-{E'_e}^4} }  \int _0^\infty dT 
T^{2-{\frac{D-1}{2}}} \int _0^\infty d\alpha_2 \sum_{n  \in {\mathbb Z } }\cr
\cr
\cr 
 && \times \, e^{-T\bigl( |t'|\alpha_2 (1-\alpha_2 )    +  \alpha_2 (n_3+n_1)^2/R^2  + n^2/R^2  + 2  \alpha_2 n(n_3+n_1) /R^2 \bigr)    } \;.
 \end{eqnarray}
We now wish to transform the amplitude to impact parameter space to check that the result is the square of the tree-level diagrams in impact parameter space, as we anticipate from the exponentiation of the eikonal phase \cite{Kabat:1992tb,Giddings:2010pp,Akhoury:2013yua}. Given that on ${\mathbb R}^{1,D-2} \times S^1  $,  the transverse momentum $q_{\perp}  = ({q_\perp}^{'}, (n_1+n_3)/R )$, with $|t'| = {q_\perp'}^2 $ in the high energy limit, we find that the impact space expression for the one-loop amplitude ${ \tilde  { \cal A}}_1 (D-1, R  ) $ is given by, 
\begin{equation}
i { \tilde  { \cal A}}_1 (D-1, R  ) = \frac{1 }{2s'}\int  \frac{d^{D-3}q_\perp'}{(2\pi )^{D-3}} e^{i q_\perp' \cdot b' }  \frac{1}{2 \pi R}  \sum_{n_3+n_1  \in {\mathbb Z } } e^{ib_s (n_3+n_1)/R}    i{ \cal A}_1 (D-1, R  ) \;,
\end{equation}
which we can see by combining \eqref{eq:defneikonal} with \eqref{eq:changeconttoKK}. The integration over the angular variables in the momentum integral $\int d^{D-3}q_\perp'$ can be carried out yielding, as usual, Bessel J functions,
\begin{eqnarray}
&& i { \tilde  { \cal A}}_1 (D-1, R  )  = -\frac{4 \kappa^4 (s'^2)^2 }{(2 s')^2} \frac{\pi ^{\frac{D+1}{2} }}{(2\pi)^{D-1}} \frac{1}{ (2\pi)^{\frac{D-3}{2}}} (b')^{\frac{5-D}{2}}  \frac{1}{(2 \pi R)^2} \int _0^\infty dT  \,  T^{\frac{5-D}{2}}  \int_0^1 d \alpha_2 \sum_{m, n \in {\mathbb Z } } \nonumber \\
&& \times  e^{im b_s/R} e^{-in b_s/R}  \left(  \int_0^\infty dq_\perp'    e^{-T q_\perp'^2 \alpha_2(1-\alpha_2)}  {( q_\perp' )}^{\frac{D-3}{2}}   J_{\frac{D-5}{2} }(q_\perp' b')  \right) e^{-T( \alpha_2 \frac{m^2}{R^2} +(1-\alpha_2 )\frac{n^2}{R^2} )}  \;.
\end{eqnarray}
The integration over the remaining momentum magnitude, $ q_\perp' $, can be carried out using the identity \cite{gradshteyn1996table},
\begin{equation}
\int_0^\infty  e^{-\alpha x^2 } x^{\nu +1} J_\nu( \beta x) dx = \frac{\beta^\nu }{(2 \alpha )^{\nu +1 } } \, e^{ - \beta^2/4\alpha} \;,
\end{equation}
where in our case we have $ \alpha  = T \alpha_2(1-\alpha_2 ) , \,  \nu = \frac{D-5}{2},  \, \beta = b'$. We then find,
\begin{eqnarray}
&&i { \tilde  { \cal A}}_1 (D-1, R  )   =  -\frac{\pi^\frac{D+1}{2} }{ {(2\pi)}^\frac{3D-5}{2}}  \frac{1}{(2s')^2}  \frac{4 \kappa^4 (s'^2)^2 }{(2 \pi R)^2}  \sum_{m, n \in {\mathbb Z }} e^{im b_s/R} e^{-in b_s/R}  \int_0^1  d\alpha_2 \int_0^\infty d {\hat T}  \cr \nonumber
  \cr 
  \cr 
  &&\times \,  {({\hat T})}^{\frac{5-D}{2}} \frac{1}{(2 {\hat T} )^\frac{D-3}{2}  } \frac{1}{\bigl( \alpha_2 (1- \alpha_2) \bigr)^{\frac{7-D}{2}}}  \,  e^{-b'^2/4 {\hat T} } \, e^{ -\frac{\hat T}{R^2} ( \frac{m^2}{1-\alpha_2 }  + \frac{n^2}{ \alpha_2 } ) } \;.
\end{eqnarray}
To perform the integration over $\alpha_2$ it is convenient to make the change of variables, $u = \frac{\alpha_2}{ 1- \alpha_2}$, leading to,
\be
I_1 = \int_0^1  d\alpha_2  \frac{ e^{ -\frac{\hat T}{R^2} ( \frac{m^2}{1-\alpha_2 }  + \frac{n^2}{\alpha_2} ) } }{\bigl( \alpha_2 (1- \alpha_2 ) \bigr)^{\frac{7-D}{2} }  } = e^{ -\frac{\hat T}{R^2} (m^2 +n^2) } \int_0^\infty du \frac{(1+u)^{5-D}}{u^{\frac{7-D}{2}}} 
e^{ -\frac{\hat T}{R^2} ( m^2 u + \frac{n^2}{u} ) } \;.
\ee
From \cite{gradshteyn1996table} we find the following integral identity,
\be
\int_0^\infty x^{\nu - 1} e^{- \beta/4x -\gamma x } \, dx = {\left( \frac{\beta}{\gamma} \right)}^{\frac{\nu}{2}} \, K_\nu (\sqrt{\beta \gamma} \,  ) \;.
\ee
Although the remaining integration over $u$ that we need to perform is not quite in this form for arbitrary $D$ it is for $D=5$. So taking $D=5$ we find, 
\be 
I_1 = K_0\left(\sqrt{\frac{2 n m {\hat T}}{R^2} } \, \right) \;.
\ee
Putting all this together we find that the one-loop amplitude for $D=5$ becomes,
\ba
i { \tilde  { \cal A}}_1 (D-1, R  )|_{D=5}   &=&  -\frac{\pi^3 }{ {(2\pi)}^5}  \frac{1}{(2s')^2}  \frac{4 \kappa^4 (s'^2)^2 }{(2 \pi R)^2}  \sum_{m, n \in {\mathbb Z }}  e^{im b_s/R -in b_s/R}  \cr
\cr
  &&\times \int_0^\infty  \frac{d {\hat T} }{2 {\hat T}} \,e^{-b'^2/4 {\hat T} -\frac{\hat T}{R^2} (m^2 + n^2 ) } K_0\left(\sqrt{\frac{2 n m {\hat T}}{R^2} }\right) \;.
\ea
An interesting integral identity involving Bessel K-functions is the so called 'duplicating' relation \cite{gradshteyn1996table},
\be
\label{Kdup}
\int_0^\infty e^{-\frac{x}{2} - \frac{1}{2x} (z^2 +\omega^2 ) }\, K_\nu \left( \frac{z\omega}{x}\right)\, \frac{dx}{x} = 2 K_\nu(z) K_\nu (\omega ) \;.
\ee 
To make use of this identity we need to perform the change of variables 
$ {\tilde T} = 1/{\hat T} $ and then $T' = \frac{{b'}^2}{2} {\tilde T}$. Performing these substitutions we then have,
\ba
i{ \tilde  { \cal A}}_1 (D-1, R  )|_{D=5} &=&  -\frac{\pi^3 }{ {(2\pi)}^5}   \frac{ \kappa^4 s'^2 }{(2 \pi R)^2} \sum_{m, n \in {\mathbb Z }}  e^{im b_s/R -in b_s/R} \nonumber \\
&& \times  \int_0^\infty  \frac{d { T'} }{{T'}} \,e^{- \frac{T'}{2}  -\frac{{b'}^2}{2 T' R^2} (m^2 + n^2 ) } K_0\left(\frac{n m{b'}^2}{T'R^2}\right) \;.
\ea
Now using the identity given in \eqref{Kdup} with $z^2 = \frac{m^2 {b'}^2}{R^2} $ and  $\omega^2 = \frac{n^2 {b'}^2}{R^2} $  (and $z = +\sqrt{z^2} = \frac{|m| b'}{R}, \omega = +\sqrt{\omega^2} = \frac{|n| b'}{R} $ ), doing so we arrive at the final form of ${ \tilde  { \cal A}}_1 (D-1, R  )|_{D=5}$,
\ba
i{ \tilde  { \cal A}}_1 (D-1, R  )|_{D=5} & = &-\frac{\pi^3 }{ {(2\pi)}^5}   \frac{ \kappa^4 s'^2 }{(2 \pi R)^2}  \sum_{m, n \in {\mathbb Z }}  e^{im b_s/R -in b_s/R}\, K_0\left(\frac{|m| b'}{R} \right) K_0\left(\frac{|n| b'}{R} \right) \cr 
\cr
\cr
 & = & -\frac{1}{2}\left( \frac{s' \kappa^2}{4 \pi}\sum_{m \in {\mathbb Z }} \frac{ e^{im b_s/R} }{2 \pi R} K_0\left(\frac{|m| b'}{R} \right)\right)^2 \cr
 \cr
 \cr 
 &=&  \frac{1}{2}\Bigl(  i { \tilde{\cal A}}_0 (D-1,R)|_{D=5}   \Bigr)^2 \;,
\ea
where we have used the ultra relativistic form of the tree-level scattering amplitude \eqref{RTLC}, for which  $s'$ is much greater than the Kaluza-Klein masses of the scattering particles. This result corroborates our intuition that the eikonal exponentiates into a phase as suggested earlier. Note that although we have taken the ultra-relativistic limit for simplicity this result extends beyond the ultra-relativistic case. It should be emphasised that the exponentiated eikonal phase ${\tilde{\cal A}}_0 (D-1,R)$ is the quantity that is directly related to classical gravitational observables as emphasised recently in \cite{Kosower:2018adc}. We shall investigate this in later sections by relating it to the deflection angles and various classical background geometries.

\section{Various Kinematic Limits of the Eikonal} \label{kinlimits}

\subsection{The Leading Eikonal in the Ultra-Relativistic Limit}

In the ultra-relativistic case where $s' \gg m_1^2, m_2^2 $ where $m_1^2 = n_1^2/R^2, m_2^2 =n_2^2/R^2 $ are the Kaluza-Klein masses,   we expect the eikonal to be related to the deflection angle or time delay of a relativistic particle moving in the background geometry corresponding to the Aichelburg-Sexl (A-S) shock-wave metric \cite{Camanho:2014apa}. However since in our case one of the transverse directions is compactified on a circle, we have to reconsider the form of the A-S metric on $ {\mathbb R}^{1,D-2} \times S^1$.

First lets review the non-compact case. In ${\mathbb R}^{1,D-1} $
the form of the A-S metric is, 

\be 
ds^2 = dudv + h(u, x^i)du^2 + \sum_{i=1}^{D-2} \, (dx^i)^2 \;,
\ee 
where $u, v$ are the usual light-cone coordinates and $x^i$, $i=1,\ldots,D-2$ are the flat transverse coordinates. For the non-compact case we will follow the discussion in \cite{Camanho:2014apa} to make contact with the eikonal. The stress-energy that sources the A-S metric is due to a relativistic particle moving in the $v$-direction carrying momentum $-P_u$ ($P_u < 0 $  producing a shock-wave at $u=0$),
\be
T_{uu} = -P_u \delta(u) \delta^{D-2}(x^i) \;.
\ee
The Einstein equations reduce to 
\be 
\nabla_{\perp}^2 h(u, x^i) = -16 \pi G_D |P_u| \delta(u) \delta^{D-2}(x^i) \;.
\ee
The function $ h(u, x^i) $ is clearly related to the Green functions of the Laplace operator on the flat transverse space, so the general solution is (for $D>4 $),
 \be
 h(u, x^i ) =  \frac{4 \Gamma \bigl( \frac{D-4}{2} \bigr)}{\pi^{\frac{D-4}{2} } } \frac{G_D |P_u| \delta(u) }{r^{D-4}} \;,
 \ee
with $r^2 = \sum_{i=1}^{D-2} \, (x^i)^2 $. In order to have continuity in $v$ as a second particle  with momentum $p_v$ moves across the shock-wave in the $u$ direction at transverse distance $r =b$ one can remove the singular term $\delta(u)$  in the metric by defining new coordinate $v_{new}$
\be
v = v_{new}  + \frac{4 \Gamma \bigl( \frac{D-4}{2} \bigr)}{\pi^{\frac{D-4}{2} } } \frac{G_D |P_u|  }{b^{D-4}}\theta(u) \;,
\ee
where $ \theta(u)$ is the usual step function. This leads to a corresponding Shapiro time delay as the particle crosses the shock wave
\be
\Delta v =  \frac{4 \Gamma \bigl( \frac{D-4}{2} \bigr)}{\pi^{\frac{D-4}{2} } } \frac{G_D |P_u|}{b^{D-4}} \;.
\ee
This result is consistent with the leading eikonal $\chi(D) $ of \eqref{TLNC} with
\be
\chi(D) = -p_v \Delta v|_{r=b} \;,
\ee 
where we identify, $s =  4p_vP_u $.

Now lets consider the case when the shock-wave is due to a particle moving in ${\mathbb R}^{1,D-2} \times S^1 $. The form of the metric becomes,
\be
\label{ASCOMP}
ds^2 = dudv + h(u, x^i, y)du^2 + \sum_{i=1}^{D-3} \, (dx^i)^2 + dy^2 \;,
\ee
where $y$ is the coordinate on $S^1$ with $y \sim y +2\pi R $. Note that 
$(x^i, y) $ are still transverse to the shock-wave. The stress-energy tensor in this case becomes,
\be
T_{uu} = -P_u \delta(u) \delta^{D-3}(x^i) \frac{1}{2\pi R}\sum_{n  \in {\mathbb Z } } e^{in y/R} \;,
\ee
where we have  given the representation of the delta function on $S^1 $ in terms of the quantised momentum modes. The Einstein equations reduce to,

\be
(\nabla_{\perp}^2 +\partial_y^2 )h(u, x^i, y) = -16 \pi G_D |P_u| \delta(u) \delta^{D-3}(x^i)\frac{1}{2\pi R}\sum_{n  \in {\mathbb Z } } e^{in y/R} \;,
\ee
where $\nabla_{\perp}^2   $ is the Laplacian on ${\mathbb R}^{D-3} $. The solution to $ h(u, x^i, y)$ is
\ba
h(u,x^i,y) &=&  16\pi |P_u| \, \frac{G_D}{2 \pi R}\,  \sum_{n  \in {\mathbb Z } } e^{in y/R} \int \frac{d^{D-3}q_\perp}{{(2\pi)}^{D-3}} \frac{ e^{i q_\perp\cdot x} }{(q_\perp^2 +n^2/R^2 ) } \delta(u) \cr
\cr
\cr
& = & 8 \frac{|P_u|}{(2\pi)^{\frac{D-5}{2} }} \, \frac{G_D}{2 \pi R} \, \frac{1}{r^{\frac{D-5}{2} } }\,  \sum_{n  \in {\mathbb Z } } e^{iy n/R}  { \left( \frac{|n|}{R} \right)^{\frac{D-5}{2} } K_{\frac{D-5}{2} } \left( \frac{|n|r}{R} \right) \delta(u) } \;,
\ea
where $K_\nu(x) $ are Bessel K functions and $r^2 = \sum_{i=1}^{D-3} \, (x^i)^2 $.

Considering a second particle moving through this geometry, now separated from the first by the impact vector $b = (b', b_s) $ i.e. the particle approaches at minimum distance $r = b'$ in ${\mathbb R}^{D-3}$ and  $y = b_s $ along the $S^1 $. The corresponding Shapiro time delay is,
 \be
 \Delta v =  8 \frac{|P_u|}{(2\pi)^{\frac{D-5}{2}} } \, \frac{G_D}{2 \pi R} \, \frac{1}{b'^{\frac{D-5}{2} } }\,  \sum_{n  \in {\mathbb Z } } e^{ib_s n/R}  {\Bigl( \frac{|n|}{R} \Bigr)^{\frac{D-5}{2} } K_{\frac{D-5}{2} } \Bigl( \frac{|n|b'}{R} \Bigr)} \;.
 \ee
This agrees with the leading eikonal expression \eqref{RTLC} in the ultra-relativistic limit  $ s' \gg m_1^2, m_2^2 $,
\be
\chi(D-1,R) =  - p_v \Delta v|_{r=b, y = b_s} \;,
\ee
where we have taken $s'= 4p_vP_u $. From the previous analysis of the eikonal $\chi(D-1,R)$ in the limits  $R \rightarrow 0 $ and  $R \rightarrow \infty $ we can see that the shock-wave metric \eqref{ASCOMP} reduces to the usual Aichelburg-Sexl form in $D-1$ and $D$ non-compact dimensions respectively, using the relation $G_D \equiv 2 \pi R G_{D-1} $.

\subsection{The Leading Eikonal in the Large Kaluza-Klein Mass Limit}

We can choose a particular set of kinematics for \eqref{eq:compeikonal} in order to reproduce and make contact with the deflection angle of a probe in the background of a Schwarzschild black hole in the ${\mathbb R}^{1,D-2}$ non-compact sector. In order to do so we will let $n_1=n_3=0$ (such that our probe has zero Kaluza-Klein mass) and $m_2^2 > s' \gg t'$ and work with $D=5$ (we also show this result for general $D$ below). In this limit we find that \eqref{eq:compeikonal} becomes,
\begin{equation}
\tilde{\mathcal{A}}_{0} \approx \frac{i \kappa_5^2 (s' - m_2^2)}{2} \int \frac{{\textrm{d}}^{2}q_{\bot}'}{(2\pi)^{3}}e^{iq_{\bot}' \cdot b'} \frac{1}{R} \frac{1}{(q_{\bot}')^2} \;.
\end{equation}
We can write the momenta of the incoming external particles explicitly as $p_1=(\mathbf{p_1'},p_{1,s})$ with $\mathbf{p_1'}=(E,\mathbf{p})$ and $p_2=(\mathbf{p_2'},p_{2,s})$ with $\mathbf{p_2'}=(m_2,0,0,0)$
We then have for $s' - m_2^2$,
\begin{eqnarray}
s' - m_2^2 = -(\mathbf{p_1'}+\mathbf{p_2'})^2 - m_2^2 = 2 E m_2 \;,
\end{eqnarray}
where we have used $E^2 - \mathbf{p}^2 = 0$. If we also use the fact that $\kappa_5^2=8 \pi G_5 = 8 \pi (2 \pi R G_4)$ we have,
\begin{eqnarray}
\tilde{\mathcal{A}}_{0} & \approx & \frac{i 8 \pi G_4}{(2 \pi)^2} (E m_2) \int {\textrm{d}}^{2}q_{\bot}'e^{iq_{\bot}' \cdot b'} \frac{1}{(q_{\bot}')^2} \nonumber \\
& = & 4 G_4 m_2 E \left( \frac{1}{D-4} - \log{b} + \ldots \right) \;.
\end{eqnarray}
We find that the equation above agrees with the results in \cite{Akhoury:2013yua}. We can now relate this to the leading order contribution to the deflection angle. Using the fact that,
\begin{equation}
\theta \approx - \frac{1}{E} \frac{\partial \chi}{\partial b} \qquad \text{where} \quad \chi = \tilde{\mathcal{A}}_{0} \;.
\end{equation}
We find for the deflection angle,
\begin{equation}
\theta = \frac{4 G_4 m_2}{b} + \ldots \;.
\end{equation}
This is the well known expression for the leading contribution deflection angle of a massless probe in the background of a Schwarzschild black hole. In fact this result is found to be universal. We find the same expression for the leading contribution to the deflection angle of a Reissner-Nordstrom black hole \cite{Eiroa:2002mk,Sereno:2003nd} as well as for the EMd black hole described in appendix \ref{angleEMDbh}. We will see in the next section that this EMd black hole is in fact the relevant black hole solution for the case we are considering here. This universality is due to the fact that no matter what source you are scattering from, the high energy limit of the Born amplitude is always the same \cite{D'Appollonio:2010ae,D'Appollonio:2013hja,Collado:2018isu}.

We can also generalise this to arbitrary spacetime dimensions $D$. In this case we find that \eqref{eq:compeikonal} becomes,

\begin{eqnarray}
\tilde{\mathcal{A}}_{0} & \approx & \frac{i 8 \pi G_{D-1}}{(2 \pi)^{D-3}} (E m_2) \int {\textrm{d}}^{D-3} q_{\bot}'e^{iq_{\bot}' \cdot b'} \frac{1}{(q_{\bot}')^2} \nonumber \\
& = & \frac{i 8 \pi G_{D-1} E m_2}{4 \pi^{\frac{D-3}{2}}} \Gamma{\left(\frac{D-5}{2} \right)} \frac{1}{b^{D-5}} \;.
\end{eqnarray}
We therefore find for the deflection angle,

\begin{eqnarray}
\theta & \approx & \frac{4 G_{D-1} m_2}{ \pi^{\frac{D-5}{2}}} \Gamma{\left(\frac{D-3}{2} \right)} \frac{1}{b^{D-4}} \nonumber \\
& = & \sqrt{\pi} \frac{\Gamma{\left(\frac{D-1}{2} \right)}}{\Gamma{\left(\frac{D-2}{2} \right)}} \left( \frac{R_s}{b} \right)^{D-4} \;,
\end{eqnarray}
where we the $D$-dimensional generalisation of the Schwarzschild radius is given by,

\begin{equation}
8 G_{D-1} m_2 = \frac{D-3}{\Gamma{\left(\frac{D-2}{2} \right)}}  \pi^{\frac{D-4}{2}} R_s^{D-4} \;.
\end{equation}
We find that this equation for the deflection angle is equivalent to the expression (D.12) found in appendix D of \cite{Collado:2018isu} (with the above modification for the Schwarzschild radius in this context). Note that here the dimensionality is shifted by 1 since the black hole resides in the $D-1$ uncompact dimensions.

\subsection{The Subleading Eikonal in the Large Kaluza-Klein Mass Limit}

In this section we find the subleading corrections to the eikonal for high energy scattering of a neutral scalar particle off a heavy Kaluza-Klein state extending the analysis on the leading eikonal in the previous section. There we showed the leading eikonal in the scalar-gravity theory compactified on ${\mathbb R}^{1,D-2} \times S^1$ reproduced the expected leading contribution to the deflection angle in the case of a EMd black hole living in $D-1$ dimensions. Which we noted was equivalent to the leading contribution of the Schwarzschild black hole result. However as is well known, massive Kaluza-Klein states also couple to the dilaton and massless gauge fields that are part of the higher-dimensional metric and as we know from \cite{D'Appollonio:2010ae,D'Appollonio:2013hja,Collado:2018isu} these interactions should start contributing at the level of the subleading eikonal.

We know that for general spacetime and world-volume dimensions there is a subleading contribution to the eikonal for a scalar scattering off of a stack of $D$p-branes \cite{Collado:2018isu}. However in the case of $D=4, p=0$ this subleading contribution vanishes. Note that in $D=4, p=0$ case the stack of $N$ $D0$-branes carry a net electric charge $N \mu_0$ and the RR field coupling to the $D0$-branes is just an abelian gauge field $C_\mu$. Since the states exchanged in the $D0$-brane scattering case are equivalent to the states exchanged between the scalar probe and heavy Kaluza-Klein state we would like to extend the results found in \cite{Collado:2018isu} to the case we are considering here.

In order to do this we will argue that both the supergravity action and the Kaluza-Klein action, \eqref{KK4D}, are equivalent in the appropriate normalisation. We will map both actions to the field normalisations used in \eqref{eq:horowitzS} so that we can also use the deflection angle results presented in appendix \ref{angleEMDbh}. From \cite{Collado:2018isu}, with $D=4$ and $p=0$, we have the supergravity action,

\begin{eqnarray}\label{eq:bulkactionSUGRA}
S_{\text{SUGRA}}= \int d^{4} x \sqrt{-g} \left[ \frac{1}{2 \kappa_4^2} R- \frac{1}{2} 
\partial_{\mu}\phi\, \partial^{\mu}\phi -
\frac{1}{4}e^{-\sqrt{6}\kappa_4 \phi }F^2_{2} \right] \text{ .}
\end{eqnarray}
Putting this action in the same normalisation as \eqref{eq:horowitzS} by changing,
\be
\kappa_4 \rightarrow \frac{1}{\sqrt{2}} \;, \quad \phi \rightarrow 2 \phi \;, \quad C^{\mu} \rightarrow 2 C^{\mu} \;,
\ee
where $C^{\mu}$ is the gauge field associated with the field strength $F_2$. We find that,
\begin{eqnarray}
\label{4dsugra}
S_{\text{SUGRA}} = \int d^{4} x \sqrt{-g} \left[ R- 2 \partial_{\mu}\phi\, \partial^{\mu}\phi - e^{-2\sqrt{3}\phi}F^2_{2} \right]  \;.
\end{eqnarray}
This is now in the same normalisation as \eqref{eq:horowitzS} and we can see that the parameter $a$ takes the value $\sqrt{3}$. Note that in this normalisation the mass and charge of the $D0$-brane become,
\be
M = \frac{N T_0}{\kappa_4} \rightarrow \sqrt{2} N T_0 \qquad Q = N \mu_0 \rightarrow 2 N \mu_0 = 2 \sqrt{2} N T_0 \;,
\ee
where we have used the fact that $\mu_p = \sqrt{2} T_p$. Hence we find a relationship between the mass and charge given by $Q=2M$.

We can now consider the Kaluza-Klein action \eqref{KK4D}. In this case we find that the action after the substitutions,
\be
\kappa_4 \rightarrow \frac{1}{\sqrt{2}} \;, \quad \sigma \rightarrow 2 \sigma \;, \quad C^{\mu} \rightarrow 2 C^{\mu} \;,
\ee
becomes,
\begin{eqnarray}
S_{\text{KK}} &=& \int d^{4} x \sqrt{-g} \biggl[ R- 2 \partial_{\mu}\sigma\, \partial^{\mu}\sigma - e^{-2\sqrt{3}\sigma}F^2_{2} \nonumber \\
&& - \frac{1}{2} \sum_{n \in {\mathbb Z }} \bigl( ( D_\mu \Phi_n )( D^\mu \Phi^{*}_n  ) +  e^{-2 \sqrt{3} \sigma} m_n^2  \Phi^{*}_n  \Phi_n \bigr) \biggr] \;.
\end{eqnarray}
We find that the massless sector of this action is equivalent to the action \eqref{eq:horowitzS} with $a=\sqrt{3}$. We can also look at what happens to the mass and charge in this case,
\be
M = \frac{n}{R} \qquad Q = \sqrt{2} \kappa_4 \frac{n}{R} \rightarrow 2 \frac{n}{R} \;,
\ee
and so again we find that $Q=2M$. This analysis suggests that we can simply use the results previously derived in the case of a $D0$-brane, for the scattering from a heavy Kaluza-Klein mode which we are considering in this paper. We have shown this equivalence explicitly for the simplest diagram with double dilaton exchange in appendix \ref{dilapp}. This then implies that we have,
\be
{\chi}^{(2)}_{\text{KK}} (R) = 0 \;,
\ee
where ${\chi}^{(2)}_{\text{KK}} (R)$ is the subleading eikonal in the Kaluza-Klein theory. We can now compare this to the deflection angle result derived in appendix \ref{angleEMDbh}. There we found that the subleading contribution to the deflection angle is given by,
\be
\phi^{(2)} = \frac{3}{8} \frac{1}{b^2} \left(\pi M^2 \left(\sqrt{1+\frac{2Q^2}{M^2}} + 9 \right) - 3 \pi Q^2 \right) \;.
\ee
Using the fact that in the normalisation used we have the relation $Q=2M$ we find that,
\be
\phi^{(2)} = 0 \;.
\ee
Which is consistent with the fact that the subleading contribution to the eikonal has been found to be zero.

\section{$D0$-$D6$ Brane Duality and Magnetically Charged Dilatonic Black Holes} \label{d0d6dual}
We have seen that the 4D  action  \eqref{4dsugra} follows from  the  dimensional  reduction of 5D Einstein gravity and has electrically charged  dilatonic black hole solutions where the deflection angle of neutral light particles agrees with the eikonal scattering of dilatons off a stack of $D0$-branes in $D=4$ not only at leading order but also subleading. The action  \eqref{4dsugra} with $a=\sqrt{3}$ is  a particular case of a more general class of D-dimensional actions considered by \cite{Duff:1993ye},
\be
\label{DLA}
I_D(d) = \frac{1}{2 \kappa_D^2 } \int d^D x \sqrt{-g} \left(  R - \frac{1}{2}  (\partial \phi )^2  - \frac{1}{2(d+1)!} e^{-\alpha(d) \phi } F_{d+1}^2  \right) \;, 
\ee
with  $\alpha^2(d) = 4 - \frac{2d {\tilde d}}{d+ {\tilde d}}  $ , $F_{d+1} = d A_d $  where the $d$-form  potential $A_d$ naturally couples to an elementary electrically charged $d$-dimensional object  i.e. $d-1 $ branes.  The parameter ${\tilde d} \equiv D- d -2 $ is the dual world-volume dimension. 

The action \eqref{4dsugra}  corresponds to the choice $D=4, d=1$  (hence $ {\tilde d} = 1$) with units  and normalisations chosen where,
\be 
\label{units}
 \kappa_4 = 1/\sqrt{2}, \quad  \phi \rightarrow 2 \phi, \quad  F_{\mu \nu} \rightarrow 2 F_{\mu \nu } \;.
\ee
The authors of \cite{Duff:1993ye} showed that \eqref{DLA} has both electrically charged `elementary'  solutions corresponding to $d-1$ branes (if we include the source action coupling the  field $A_d$ to the world-volume of the d-1 brane)  as well as solitonic magnetic solutions to the action $I_D(d) $ without sources,  interpreted as magnetic ${\tilde d}-1 $ branes. The electric and magnetic charges 
$(e_d, g_{\tilde d} ) $ satisfy Dirac quantisation condition,
\be 
e_d  g_{\tilde d} = 2\pi n \;,
\ee 
with  $ e_d = 2 \kappa_D T_d (-)^{(D-d)(d+1) }    $  and the mass per unit volume of the $d-1$ brane ${\cal M}_d =  \frac{1}{\sqrt{2} \kappa_D }  |e_d | $ 
(choosing the dilaton vev, $ \phi_0 = 0 $). The choice $D=4$, $d=1$   makes contact with the stack of $N$ $D0$-branes discussed before and the elementary solutions (see appendix \ref{angleEMDbh} for details) describe electrically charged dilatonic black holes in the extremal limit of $Q = 2M $, where in our conventions \eqref{units}, $ Q = 2 N e_1 = 2\sqrt{2} NT_0 = 2 M $ with  $Q$ being the total electric charge of the $N$ $D0$-branes and $M$ their total mass per unit volume.

In \cite{Duff:1993ye} a dual action to \eqref{DLA} was proposed in which the roles of the equations of motion and Bianchi identities are interchanged. The action has elementary solutions corresponding to magnetically charged ${\tilde d}-1 $ branes  as well as electrically charged solitonic $d-1$ branes,
\be
\label{DDLA}
I_D({\tilde d}) = \frac{1}{2 \kappa_D^2 } \int d^D x \sqrt{-g} \left(  R - \frac{1}{2}  (\partial \phi )^2  - \frac{1}{2({\tilde d}+1)!} e^{\alpha(d) \phi } {\tilde F}_{{\tilde d}+1}^2  \right) \;.
\ee
Note the sign of $\alpha(d) $ has changed sign compared to \eqref{DLA} and the dual field strength ${\tilde F}_{{\tilde d}+1} $ is defined as,
\be 
{\tilde F}_{{\tilde d}+1} =  e^{\alpha(d) \phi }*F \;.
\ee
An interesting feature of this action is that the solutions mentioned have the same metric as \eqref{DLA}, the dilaton solutions are the same but with the sign of  $\alpha(d) $ reversed and the electric/magnetic monopole solutions for the field strengths are interchanged. In the case of $D=4, {\tilde d} =1$ we get a dual action to \eqref{4dsugra},
\be
I_4({\tilde d =1}) = \frac{1}{2 \kappa_4^2 } \int d^4 x \sqrt{-g} \left(  R - \frac{1}{2}  (\partial \phi )^2  - \frac{1}{4} e^{\sqrt{3} \phi } {\tilde F_2}^2  \right) \;. \label{DSUGRA}
\ee
The explicit electrically charged solutions of \eqref{DLA} for $D=4$ and in the normalisation  \eqref{units}  are,
\ba
&&ds^2 = -\left(1+\frac{k_1}{y} \right)^{-\frac{1}{2} } dt^2 + \left(1+\frac{k_1}{y } \right)^{\frac{1}{2} } (dy^2 +y^2 d\Omega_2^2 ) \;, \nonumber \\
&&e^{2 \phi} = \left(1 +\frac{k_1}{y} \right)^{-\frac{\sqrt{3}}{2}} ;  \qquad  A_0 = - \left(1 +\frac{ k_1}{y} \right)^{-1} ; \qquad F_{0y} = \frac{ k_1}{(y+k_1)^2} \;,
\ea
where $k_1 = 2Q = 4M$. We can see this by taking the extremal limit  $r_+ = r_- $ of the solution detailed in appendix \ref{angleEMDbh} and making the coordinate redefinition $  r = y+r_+ $ where we then find the above solution  upon identifying  $k_1 = r_+ $ which then  gives $r_+ = 2Q $ and $M =r_+/4 = Q/2 $.

As was pointed out in \cite{Duff:1993ye}  the actions  \eqref{DLA} and \eqref{DSUGRA}  can be thought of as originating from the dimensional reduction of the $D=10$ actions in the case of a $D0$-brane or a $D6$-brane, and that both actions describe dual D-particles in $D=4$.  Starting  from the action \eqref{DLA}  with $D=10$ and $d=1$ we have a $D0$-brane. By standard dimensional reduction over a single dimension,  $(I_D(d), {\tilde d})  \rightarrow ( I_{D-1}(d), \tilde d-1) $, we can then dimensionally reduce from $D=10$ to $D=4$ and we  find $(I_{D=10}(d=1), {\tilde d}=7)  \rightarrow ( I_{D=4}(d=1), {\tilde d}=1) $. This is the action given by \eqref{4dsugra}.

On the other hand we can perform a different dimensional reduction, the so called `double' dimensional reduction on the dual action \eqref{DDLA} with $D=10$. Such a reduction involves directions parallel to the ${\tilde d}-1 $ brane, so we have in general dimension D,  $(I_D({\tilde d}), d)  \rightarrow ( I_{D-1}({\tilde d}), d-1) $. If we know consider a $D6$  brane in $D=10$, then double dimensional reduction yields a particle action in $D=4$, $(I_{D=10}({\tilde d}=1), d=7 )  \rightarrow ( I_{D=4}(d=1), {\tilde d}=1) $. This is the action \eqref{DSUGRA}. Thus $D0$-$D6$ brane particle/brane duality in $D=10$ reveals itself through the dual $D=4$ actions \eqref{4dsugra} and \eqref{DSUGRA}. 

As mentioned both actions have solutions (elementary or solitonic) with the same metric. This implies that both actions give the same expressions for the deflection angle of massless particles moving along geodesics in the corresponding curved geometries. In the case of a stack of $D0$-branes we saw that the eikonal for dilatonic scattering has vanishing subleading contributions in $D=4$ which is consistent with the absence of such terms in the deflection angle formula derived in appendix \ref{angleEMDbh} (when one considers the extremal case $Q=2M$). From what we have discussed above, the same should be true for scattering dilatons off a stack of $D6$-branes in $D=10$. In \cite{Collado:2018isu} it was shown that the expression for the subleading eikonal for scattering off a stack of $N$ $Dp$-branes has a factor of $\Gamma(D-p-4)$ in the denominator. This factor diverges, and so the subleading contribution vanishes, not only when $D=4$ and $p=0$ but also for $D=10$ and $p=6$. 

The corresponding extremal black holes in the case of $D6$-branes are magnetically charged with extremal values $P = 2M$, with $P$ being the magnetic charges in geometric units. These are just the dual of the extremal electrically charged $Q=2M$ black holes discussed earlier.  The general (non extremal) magnetically charged black holes are just the solutions to the dual action \eqref{DSUGRA}.

\section{Discussion} \label{disc}
In this paper we have investigated eikonal scattering in one of the simplest examples of a Kaluza-Klein theory, namely 5D Einstein gravity coupled to a real, massless 5D free scalar field, compactified to 4D on a circle. Despite this there is a richness to the 4D theory since it contains, in the massless sector, a massless scalar originating from the 5D massless scalar and a graviton, $U(1)$  gauge file and a dilaton, which originate from the 5D graviton. The massive sector contains a Kaluza-Klein tower of massive scalars charged with respect to the gauge field whose mass and charge satisfy the relation $Q= 2M $ in appropriate units. Their masses and charges have the usual interpretation of quantized momentum states moving around the circle. In considering $2 \rightarrow 2$ scattering of these massive 4D scalars there are both elastic and inelastic processes involved. Thus from the 4D point of view it is quite a complex system. It has already been previously shown that for simple scalar gravity theories in non-compact dimensions, contributions with a higher number of gravitons exchanged exponentiate into a phase \cite{Kabat:1992tb,Giddings:2010pp,Akhoury:2013yua} which we have briefly reviewed here. Thus our model is an extension of this case. The proof that the eikonal exponentiates to all orders in our model is something that would be interesting to investigate. We have seen that the one-loop contribution (double graviton exchange) is indeed the square of the tree-level contribution (single graviton exchange) as implied by the exponentiation of the eikonal phase. One may intuitively expect that the exponentiation holds at all orders since in the limit of infinite or zero compactification radius $R$ we recover the 5D or 4D Einstein-scalar theories, for which we know the eikonal phase exponentiates. But explicitly proving this even at the one-loop level, as we have shown, is rather non-trivial. 

An interesting test of our expression for the eikonal phase in the compactified case were the comparison and agreement with deflection angle calculations in the various corresponding background geometries. We found agreement in the ultra-relativistic limit where the corresponding geometry was the compactified version of the Aichelburg-Sexl shock wave metric. We also considered the heavy Kaluza-Klein mass limit where we found agreement with the deflection angle calculated from a Einstein-Maxwell-dilaton black hole. This was as one might have expected because the heavy Kaluza-Klein modes are also charged under the $U(1)$ in 4D and also couple to the dilaton.  

There has been recent interest in investigating binary black hole systems in the post-Minkowskian approximation \cite{Damour:2016gwp,Damour:2017zjx} as these results could provide useful theoretical modelling of  gravitational waves detected by LIGO \cite{Abbott:2016blz}. Recent work \cite{Khalil:2018aaj} has investigated the particular case of binary Einstein-Maxwell-dilaton black holes in the post-Newtonian approximation. We believe that our result for the eikonal in Kaluza-Klein theories,
\begin{equation}
\chi (D-1, R)  = \frac{\kappa^2 }{2} \frac{s' - m_1^2 -m_2^2 - 2 m_1 m_2  }{(2\pi)^{\frac{D-1}{2}} }\frac{1}{ {(b')}^{\frac{D-5}{2} }}  \frac{1}{R}  \sum_{n  \in {\mathbb Z } } e^{ib_s n/R}  {\biggl( \frac{|n|}{R} \biggr)^{\frac{D-5}{2} } K_{\frac{D-5}{2} } \biggl( \frac{|n|b'}{R} \biggr) \;,
}\end{equation}
could be relevant in the study of such binary systems in the first post-Minkowskian approximation.

The massless sector of the Kaluza-Klein action given by \eqref{KK4D} supported the well known Einstein-Maxwell-dilaton black hole solutions whose extremal $Q=2M$ limit gave deflection angles consistent with the eikonal calculation in the limit where one of the 2 scattering massive Kaluza-Klein scalar is taken to be very heavy. We also saw that there was the natural dual action given in \eqref{DDLA} which has dilatonic magnetic black hole solutions. In the electric case we could understand the heavy electrically charged Kaluza-Klein mode as sourcing the background Einstein-Maxwell-dilaton black hole geometry felt by the massless scattering particle. In the magnetic case it is not clear, from a field theory point of view, what the dual 'sources' to the massive electrically charged Kaluza-Klein scalars are. However in the context of string theory, scattering dilatons off stacks of $D0$- or $D6$-branes, we saw that the situation was clearer. Dimensionally reducing both cases from $D=10$ to $D=4$ gives rise to the dual actions \eqref{4dsugra}  and \eqref{DSUGRA} respectively \cite{Duff:1993ye}. The electric/magnetically charged point particles in $D=4$ are just the $D0$-branes and the wrapped $D6$-branes respectively. 

Apart from the purely magnetic or electric dilatonic black hole solutions we have discussed in this paper, there exist explicit solutions describing more general dyonic black holes carrying both kinds of charges and which also carry angular momentum \cite{Dhar:1998ip,Larsen:1999pp}. It is interesting to speculate that the deflection angle calculation in these geometries should correspond to eikonal scattering of dilatons off a stack of bounds states of $D0$ and $D6$-branes. Bound states of $D1$-$D5$ branes in IIB string theory and eikonal scattering have been investigated in \cite{Giusto:2009qq,Bianchi:2017sds}. The $D0$-$D6$ brane bound state systems are more involved as a priori they are not stable unless one turns on a background  $B_{\mu \nu}$ field \cite{Witten:2000mf}. Nevertheless if one can derive suitable Feynman rules for the coupling of gravitons, RR fields and dilatons to such bound state sources, the eikonal computations could be carried out along the lines described in \cite{Collado:2018isu}. Since the $D0$-$D6$ bound states carry orbital angular momentum it could provide yet another interesting connection between scattering amplitudes on the one hand, and gravitational waves sourced by rotating binary black hole systems \cite{Bini:2017xzy,Vines:2017hyw,Bini:2018ywr,Guevara:2018wpp}.

\section*{Acknowledgements}
We would like to thank  R. Russo for useful comments and suggestions. This work was supported in part by the Science and Technology Facilities Council (STFC) Consolidated Grant ST/L000415/1 {\it String theory, gauge theory \& duality}. AKC is supported by an STFC studentship.

\appendix

\section{Deflection Angle for Einstein-Maxwell-dilaton Black Hole} \label{angleEMDbh}

We can consider black hole solutions in a theory with a graviton, dilaton and photon with an action given by \cite{Horne:1992zy},
\be
\label{eq:horowitzS}
S = \int d^4 x \sqrt{-g} \left(R - 2(\nabla \Phi)^2 - e^{-2 \alpha \Phi}F^2 \right) \;,
\ee
where units have been taken such that $\kappa=1/\sqrt{2}$ and the spacetime dimension has been taken to be $D=4$. The solution describing static, spherically-symmetric black holes in this theory is given by,
\ba
\label{genericmetric}
ds^2 &=& -A(r) dt^2 + B(r) dr^2 + C(r) d \Omega^2  \\
e^{2 \phi} &=&  \left( 1- \frac{r_-}{r} \right)^{\sqrt{1 -\gamma^2}} \\
F_{tr}  &=& \frac{Q}{r^2}  \;,
\ea
where we have,
\begin{eqnarray}
A(r) &=& \left( 1 - \frac{r_{+}}{r} \right)\left( 1 - \frac{r_{-}}{r} \right)^{\gamma} \label{metriccomp1} \\
B(r) &=& A^{-1}(r) \label{metriccomp2} \\
C(r) &=& r^2 \left( 1 - \frac{r_{-}}{r} \right)^{1-\gamma} \label{metriccomp3} \;,
\end{eqnarray}
with $\gamma={\frac{1-\alpha^2}{1+\alpha^2}}$ and the various symbols introduced are related to the mass and charge of the black hole via,
\be
2M = r_{+} + \gamma r_{-} \qquad 2 Q^2 = (1+\gamma)r_{+}r_{-} \;.
\ee
Inverting the equations above we find that we can write,
\be
r_{+} = M + M \sqrt{1-\frac{2 \gamma}{1+\gamma}\left(\frac{Q}{M} \right)^2 } \qquad r_{-} = \frac{M}{\gamma} - \frac{M}{\gamma} \sqrt{1-\frac{2 \gamma}{1+\gamma}\left(\frac{Q}{M} \right)^2 } \;.
\ee
Below we will consider the deflection angle for general parameter $\gamma$, however it is worth noting at this stage that the above solution  takes an interesting form in the specific case when $\alpha = \sqrt{3} $ and we take the extremal limit  $r_- = r_+ $,  or correspondingly  $Q = 2M$. The metric becomes
\be
ds^2 = -\left( 1 + \frac{r_+}{r} \right)^{\frac{1}{2}} dt^2 + \left( 1 + \frac{r_+}{r} \right)^{-\frac{1}{2}} dr^2 + r^2\left( 1 + \frac{r_+}{r} \right)^{\frac{3}{2} } d\Omega^2
\ee
For any generic metric in the form of \eqref{genericmetric} we can write the deflection angle, $\phi$, of a massless probe scattering from the black hole as,
\be
\phi + \pi = 2 \int_{r_0}^{\infty} dr \left( \frac{B(r)}{C(r)} \right)^{\frac{1}{2}} \left( \frac{C(r)A(r_0)}{C(r_0)A(r)} - 1 \right)^{-\frac{1}{2}} \;, \label{gendeflang}
\ee
where $r_0$ is the point of closest approach and is related to the impact parameter by,
\be
b(r_0) = \sqrt{\frac{C(r_0)}{A(r_0)}} \;. \label{breltor0}
\ee
Substituting the quantities given by \eqref{metriccomp1}--\eqref{metriccomp3} into \eqref{gendeflang} we find that in the case we are considering here we can write the deflection angle as,
\be
\phi + \pi = 2 \int_{r_0}^{\infty} dr \frac{1}{r^2 \left( 1 - \frac{r_{-}}{r} \right)^{1-\gamma}} \left[ \frac{\left( 1 - \frac{r_{+}}{r_0} \right) \left( 1 - \frac{r_{-}}{r_0} \right)^{\gamma}}{r_0^2 \left( 1 - \frac{r_{-}}{r_0} \right)^{1-\gamma}} - \frac{\left( 1 - \frac{r_{+}}{r} \right) \left( 1 - \frac{r_{-}}{r} \right)^{\gamma}}{r^2 \left( 1 - \frac{r_{-}}{r} \right)^{1-\gamma}} \right] \;.
\ee
Using the substitution $u=r_0/r$ we find,
\be
\phi + \pi = 2 \int_{0}^{1} du \left[\left( 1 - \frac{r_{+}}{r_0} \right) \left( 1 - \frac{r_{-} u}{r_0} \right)\left( \frac{r_0 - r_{-}}{r_0 - r_{-}u} \right)^{2\gamma-1} - u^2 \left( 1 - \frac{r_{+} u}{r_0} \right)\left( 1 - \frac{r_{-}u}{r_0} \right)     \right]^{-\frac{1}{2}} \;.
\ee
We want to solve this integral as an expansion in $1/r_0$, so expanding the integrand in powers of $1/r_0$ up to second order,
\be
\phi + \pi = 2 \left( \int_{0}^{1} du \frac{1}{\sqrt{1-u^2}} + \int_{0}^{1} du A_1(u) \left(\frac{1}{r_0}\right) + \int_{0}^{1} du A_2(u) \left(\frac{1}{r_0}\right)^2 + \ldots \right) \;,
\ee
with
\begin{eqnarray}
\int_{0}^{1} du A_1(u) &=& = 2 M \\
\int_{0}^{1} du A_2(u) &=& \sqrt{\frac{(\gamma+1)M^2 - 2\gamma Q^2}{(\gamma+1)M^2}} M^2 \left(\frac{\pi}{16 \gamma^2} - \frac{1}{\gamma} - \frac{\pi}{16} + 1 \right) \nonumber \\
&& + \frac{1}{16 \gamma^2} [(\pi (31\gamma^2-1)+16(1-3\gamma)\gamma) M^2 + \pi (1-7\gamma)\gamma Q^2 ] \;,
\end{eqnarray}
where we have written the results for the integrals in terms of the black hole mass and charge. We therefore have,
\begin{eqnarray}
\phi & \approx & \frac{4M}{r_0} + \frac{2}{r_0^2} \left( \sqrt{\frac{(\gamma+1)M^2 - 2\gamma Q^2}{(\gamma+1)M^2}} M^2 \left(\frac{\pi}{16 \gamma^2} - \frac{1}{\gamma} - \frac{\pi}{16} + 1 \right) \right. \nonumber \\
&& \left. + \frac{1}{16 \gamma^2} [(\pi (31\gamma^2-1)+16(1-3\gamma)\gamma) M^2 + \pi (1-7\gamma)\gamma Q^2 ] \right) \;.
\end{eqnarray}
We can check this result by computing the $\gamma=1$ case which corresponds to a Reissner–Nordstrom black hole. Choosing $\gamma=1$ and approximating $r_0$ to first order in the impact parameter using \eqref{breltor0}, $r_0 \approx b(1- M/b)$, we find,
\be
\phi \approx \frac{4M}{b} + \frac{15 \pi}{4} \frac{M^2}{b^2} - \frac{3 \pi}{4} \frac{Q^2}{b^2} \;,
\ee
as expected for a Reissner–Nordstrom black hole \cite{Eiroa:2002mk,Sereno:2003nd}.

We can also compute the relevant deflection angle in the Kaluza-Klein case by setting  $\gamma=-1/2$ (equivalently $\alpha=\sqrt{3}$). In this case we find that the deflection angle is given by,
\be
\phi \approx \frac{4M}{r_0} + \frac{1}{r_0^2} \left(\frac{M^2}{8} \left(3 \pi \sqrt{1+\frac{2Q^2}{M^2}} + 27 \pi + 48 \sqrt{1+\frac{2Q^2}{M^2}} - 80 \right) - \frac{9\pi Q^2}{8}  \right) \;.
\ee
We can again write this in terms of impact parameter space by approximating the solution to \eqref{breltor0}. We find up to first order in $b$,
\be
r_0 \approx b + \frac{M}{2} \left( 3 \sqrt{1+\frac{2Q^2}{M^2}} -5 \right) \;.
\ee
Substituting this into our previous equation for the deflection angle and expanding in powers of $1/b$ we find,
\be
\phi \approx \frac{4M}{b} + \frac{3}{8} \frac{1}{b^2} \left(\pi M^2 \left(\sqrt{1+\frac{2Q^2}{M^2}} + 9 \right) - 3 \pi Q^2 \right) \;.
\ee

\section{Double Dilaton Exchange on ${\mathbb R}^{1,3} \times S^1$} \label{dilapp}
In this appendix we will explicitly calculate the subleading contribution coming from double dilaton exchange on ${\mathbb R}^{1,3} \times S^1$ from the 4-dimensional perspective where the 5-dimensional graviton contribution is split into a 4-dimensional graviton, gauge field and dilaton field contribution. We derive the dilaton-scalar vertex from the action \eqref{KK4D} and find this to be equal to $-i \sqrt{6} \kappa_4 m^2$ as before. 

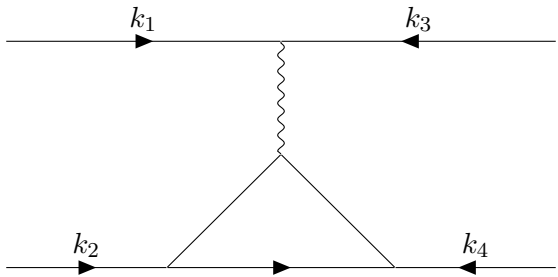
\begin{figure}[h]
  \centering
  \begin{tikzpicture}[scale=1.5]
		\begin{feynman}

			\vertex (c) at (-2.5, 0) {};
			\vertex (d) at ( 2.5, 0) {};
			\vertex (x) at (-2.5, -2) {};
			\vertex (y) at ( 2.5, -2) {};
			\vertex[circle,inner sep=0pt,minimum size=0pt] (e) at (0, -1) {};
			\vertex[circle,inner sep=0pt,minimum size=0pt] (m) at (0, 0) {};
			\vertex[circle,inner sep=0pt,minimum size=0pt] (a) at (-1, -2) {};
			\vertex[circle,inner sep=0pt,minimum size=0pt] (b) at (1, -2) {}; 
     
			\diagram* {
			(c) -- [fermion,edge label=$k_1$] (m) -- [anti fermion,edge label=$k_3$] (d),
			(m) -- [boson] (e),
			(a) -- (e),
			(b) -- (e),
			(x) -- [fermion,edge label=$k_2$] (a) -- [fermion] (b) -- [anti fermion,edge label=$k_4$] (y),
			};
		\end{feynman}
  \end{tikzpicture}
  \caption{Feynman diagram representation for one-loop scalar scattering with two dilaton exchanges. The solid lines represent scalar states and the wavy lines represent gravitons. Note that the internal solid lines represent massless dilaton states.}
  \label{fig:dildilloop}
\end{figure}

We want to calculate the diagram in figure \ref{fig:dildilloop}. Taking the probe scalar to be massless, i.e. $n_1=-n_3=0$, and the mass of the Kaluza-Klein source, $m$, to be large, we find that,
\begin{eqnarray}
\mathcal{A}^{\text{dil}}_{\text{KK}} &=&  3 \kappa_4^4 m^4 \left(\frac{1}{\sqrt{2m}} \right)^2 \int \frac{d^4 k}{(2\pi)^4} \frac{1}{q^2} \frac{1}{k^2} \frac{1}{(k-q)^2} \frac{1}{(k_2+k)^2+ m^2} \bigl( 8 (k \cdot k_1)(k \cdot k_3)  \nonumber \\
&& \;\; - 4 k^2 (k_1 \cdot k_3) \bigr) \nonumber \\
&=& \frac{3 \kappa_4^4 m^3}{2}\frac{1}{q^2} \left( 8 k_{1\mu}k_{3\nu} \mathcal{I}_{3}^{\mu \nu} + \ldots \right) \;,
\end{eqnarray}
where $k$ is the momentum in the loop, we have divided by $\sqrt{2m}$ for each external massive particle and in the second line the ellipses refer to contributions which are subleading. We can show that in the large mass limit the integral above, $2 \mathcal{I}_{3}^{\mu \nu} \rightarrow \frac{i}{2m} \mathcal{I}_{2}^{\mu \nu}$. Substituting this we find, 
\begin{equation}
\mathcal{A}^{\text{dil}}_{\text{KK}} \approx \frac{3 \kappa_4^4 m^3}{8} \frac{1}{q^2} 8 k_{1\mu}k_{3\nu} \mathcal{I}_{2}^{\mu \nu} \;,
\end{equation}
where we have neglected terms that are subleading in energy or mass. This is equivalent to the result found for the $D0$-brane in 4-dimensions in \cite{Collado:2018isu} where we account for the change of labelling $2 \leftrightarrow 3$ and take the $D0$-brane mass to be $m=N T_{p=0} / \kappa_4$.


\providecommand{\href}[2]{#2}\begingroup\raggedright\endgroup

\begin{thebibliography}{10}

\bibitem{'tHooft:1987rb}
G.~'t~Hooft, \emph{{Graviton Dominance in Ultrahigh-Energy Scattering}},
  \href{https://doi.org/10.1016/0370-2693(87)90159-6}{\emph{Phys. Lett.}
  {\bfseries B198} (1987) 61}.

\bibitem{Muzinich:1987in}
I.~J. Muzinich and M.~Soldate, \emph{{High-Energy Unitarity of Gravitation and
  Strings}}, \href{https://doi.org/10.1103/PhysRevD.37.359}{\emph{Phys. Rev.}
  {\bfseries D37} (1988) 359}.

\bibitem{Kabat:1992tb}
D.~N. Kabat and M.~Ortiz, \emph{{Eikonal quantum gravity and Planckian
  scattering}}, \href{https://doi.org/10.1016/0550-3213(92)90627-N}{\emph{Nucl.
  Phys.} {\bfseries B388} (1992) 570}
  [\href{https://arxiv.org/abs/hep-th/9203082}{{\ttfamily hep-th/9203082}}].

\bibitem{Giddings:2010pp}
S.~B. Giddings, M.~Schmidt-Sommerfeld and J.~R. Andersen, \emph{{High energy
  scattering in gravity and supergravity}},
  \href{https://doi.org/10.1103/PhysRevD.82.104022}{\emph{Phys. Rev.}
  {\bfseries D82} (2010) 104022}
  [\href{https://arxiv.org/abs/1005.5408}{{\ttfamily 1005.5408}}].

\bibitem{Akhoury:2013yua}
R.~Akhoury, R.~Saotome and G.~Sterman, \emph{{High Energy Scattering in
  Perturbative Quantum Gravity at Next to Leading Power}},
  \href{https://arxiv.org/abs/1308.5204}{{\ttfamily 1308.5204}}.

\bibitem{Melville:2013qca}
S.~Melville, S.~G. Naculich, H.~J. Schnitzer and C.~D. White, \emph{{Wilson
  line approach to gravity in the high energy limit}},
  \href{https://doi.org/10.1103/PhysRevD.89.025009}{\emph{Phys. Rev.}
  {\bfseries D89} (2014) 025009}
  [\href{https://arxiv.org/abs/1306.6019}{{\ttfamily 1306.6019}}].

\bibitem{Camanho:2014apa}
X.~O. Camanho, J.~D. Edelstein, J.~Maldacena and A.~Zhiboedov, \emph{{Causality
  Constraints on Corrections to the Graviton Three-Point Coupling}},
  \href{https://doi.org/10.1007/JHEP02(2016)020}{\emph{JHEP} {\bfseries 02}
  (2016) 020} [\href{https://arxiv.org/abs/1407.5597}{{\ttfamily 1407.5597}}].

\bibitem{Amati:1987wq}
D.~Amati, M.~Ciafaloni and G.~Veneziano, \emph{{Superstring Collisions at
  Planckian Energies}},
  \href{https://doi.org/10.1016/0370-2693(87)90346-7}{\emph{Phys. Lett.}
  {\bfseries B197} (1987) 81}.

\bibitem{D'Appollonio:2010ae}
G.~D'Appollonio, P.~Di~Vecchia, R.~Russo and G.~Veneziano, \emph{{High-energy
  string-brane scattering: Leading eikonal and beyond}},
  \href{https://doi.org/10.1007/JHEP11(2010)100}{\emph{JHEP} {\bfseries 1011}
  (2010) 100} [\href{https://arxiv.org/abs/1008.4773}{{\ttfamily 1008.4773}}].

\bibitem{D'Appollonio:2013hja}
G.~D'Appollonio, P.~Vecchia, R.~Russo and G.~Veneziano, \emph{{Microscopic
  unitary description of tidal excitations in high-energy string-brane
  collisions}}, \href{https://doi.org/10.1007/JHEP11(2013)126}{\emph{JHEP}
  {\bfseries 1311} (2013) 126}
  [\href{https://arxiv.org/abs/1310.1254}{{\ttfamily 1310.1254}}].

\bibitem{Collado:2018isu}
A.~K. Collado, P.~Di~Vecchia, R.~Russo and S.~Thomas, \emph{{The subleading
  eikonal in supergravity theories}},
  \href{https://doi.org/10.1007/JHEP10(2018)038}{\emph{JHEP} {\bfseries 10}
  (2018) 038} [\href{https://arxiv.org/abs/1807.04588}{{\ttfamily
  1807.04588}}].

\bibitem{Bjerrum-Bohr:2016hpa}
N.~E.~J. Bjerrum-Bohr, J.~F. Donoghue, B.~R. Holstein, L.~Plante and
  P.~Vanhove, \emph{{Light-like Scattering in Quantum Gravity}},
  \href{https://doi.org/10.1007/JHEP11(2016)117}{\emph{JHEP} {\bfseries 11}
  (2016) 117} [\href{https://arxiv.org/abs/1609.07477}{{\ttfamily
  1609.07477}}].

\bibitem{Bjerrum-Bohr:2018xdl}
N.~E.~J. Bjerrum-Bohr, P.~H. Damgaard, G.~Festuccia, L.~Planté and P.~Vanhove,
  \emph{{General Relativity from Scattering Amplitudes}},
  \href{https://arxiv.org/abs/1806.04920}{{\ttfamily 1806.04920}}.

\bibitem{Emparan:2001ce}
R.~Emparan, \emph{{Exact gravitational shock waves and Planckian scattering on
  branes}}, \href{https://doi.org/10.1103/PhysRevD.64.024025}{\emph{Phys. Rev.}
  {\bfseries D64} (2001) 024025}
  [\href{https://arxiv.org/abs/hep-th/0104009}{{\ttfamily hep-th/0104009}}].

\bibitem{Horne:1992zy}
J.~H. Horne and G.~T. Horowitz, \emph{{Rotating dilaton black holes}},
  \href{https://doi.org/10.1103/PhysRevD.46.1340}{\emph{Phys. Rev.} {\bfseries
  D46} (1992) 1340} [\href{https://arxiv.org/abs/hep-th/9203083}{{\ttfamily
  hep-th/9203083}}].

\bibitem{Kaluza:1921tu}
T.~Kaluza, \emph{{Zum Unitätsproblem der Physik}}, {\emph{Sitzungsber. Preuss.
  Akad. Wiss. Berlin (Math. Phys.)} {\bfseries 1921} (1921) 966}
  [\href{https://arxiv.org/abs/1803.08616}{{\ttfamily 1803.08616}}].

\bibitem{Overduin:1998pn}
J.~M. Overduin and P.~S. Wesson, \emph{{Kaluza-Klein gravity}},
  \href{https://doi.org/10.1016/S0370-1573(96)00046-4}{\emph{Phys. Rept.}
  {\bfseries 283} (1997) 303}
  [\href{https://arxiv.org/abs/gr-qc/9805018}{{\ttfamily gr-qc/9805018}}].

\bibitem{Blau}
M.~Blau, \emph{{Lecture Notes on General Relativity}}, (2018) 
\url{http://www.blau.itp.unibe.ch/Lecturenotes.html}.

\bibitem{Giddings:2011xs}
S.~B. Giddings, \emph{{The gravitational S-matrix: Erice lectures}},
  \href{https://doi.org/10.1142/9789814522489_0005}{\emph{Subnucl. Ser.}
  {\bfseries 48} (2013) 93} [\href{https://arxiv.org/abs/1105.2036}{{\ttfamily
  1105.2036}}].

\bibitem{gradshteyn1996table}
I.~Gradshteyn, A.~Jeffrey and I.~Ryzhik, \emph{Table of Integrals, Series, and
  Products}. Academic Press, 1996.

\bibitem{Kosower:2018adc}
D.~A. Kosower, B.~Maybee and D.~O'Connell, \emph{{Amplitudes, Observables, and
  Classical Scattering}},  \href{https://arxiv.org/abs/1811.10950}{{\ttfamily
  1811.10950}}.

\bibitem{Eiroa:2002mk}
E.~F. Eiroa, G.~E. Romero and D.~F. Torres, \emph{{Reissner-Nordstrom black
  hole lensing}}, \href{https://doi.org/10.1103/PhysRevD.66.024010}{\emph{Phys.
  Rev.} {\bfseries D66} (2002) 024010}
  [\href{https://arxiv.org/abs/gr-qc/0203049}{{\ttfamily gr-qc/0203049}}].

\bibitem{Sereno:2003nd}
M.~Sereno, \emph{{Weak field limit of Reissner-Nordstrom black hole lensing}},
  \href{https://doi.org/10.1103/PhysRevD.69.023002}{\emph{Phys. Rev.}
  {\bfseries D69} (2004) 023002}
  [\href{https://arxiv.org/abs/gr-qc/0310063}{{\ttfamily gr-qc/0310063}}].

\bibitem{Duff:1993ye}
M.~J. Duff and J.~X. Lu, \emph{{Black and super p-branes in diverse
  dimensions}}, \href{https://doi.org/10.1016/0550-3213(94)90586-X}{\emph{Nucl.
  Phys.} {\bfseries B416} (1994) 301}
  [\href{https://arxiv.org/abs/hep-th/9306052}{{\ttfamily hep-th/9306052}}].

\bibitem{Damour:2016gwp}
T.~Damour, \emph{{Gravitational scattering, post-Minkowskian approximation and
  Effective One-Body theory}},
  \href{https://doi.org/10.1103/PhysRevD.94.104015}{\emph{Phys. Rev.}
  {\bfseries D94} (2016) 104015}
  [\href{https://arxiv.org/abs/1609.00354}{{\ttfamily 1609.00354}}].

\bibitem{Damour:2017zjx}
T.~Damour, \emph{{High-energy gravitational scattering and the general
  relativistic two-body problem}},
  \href{https://doi.org/10.1103/PhysRevD.97.044038}{\emph{Phys. Rev.}
  {\bfseries D97} (2018) 044038}
  [\href{https://arxiv.org/abs/1710.10599}{{\ttfamily 1710.10599}}].

\bibitem{Abbott:2016blz}
{\scshape LIGO Scientific, Virgo} collaboration, B.~P. Abbott et~al.,
  \emph{{Observation of Gravitational Waves from a Binary Black Hole Merger}},
  \href{https://doi.org/10.1103/PhysRevLett.116.061102}{\emph{Phys. Rev. Lett.}
  {\bfseries 116} (2016) 061102}
  [\href{https://arxiv.org/abs/1602.03837}{{\ttfamily 1602.03837}}].

\bibitem{Khalil:2018aaj}
M.~Khalil, N.~Sennett, J.~Steinhoff, J.~Vines and A.~Buonanno, \emph{{Hairy
  binary black holes in Einstein-Maxwell-dilaton theory and their
  effective-one-body description}},
  \href{https://doi.org/10.1103/PhysRevD.98.104010}{\emph{Phys. Rev.}
  {\bfseries D98} (2018) 104010}
  [\href{https://arxiv.org/abs/1809.03109}{{\ttfamily 1809.03109}}].

\bibitem{Dhar:1998ip}
A.~Dhar and G.~Mandal, \emph{{Probing four-dimensional nonsupersymmetric black
  holes carrying D0-brane and D6-brane charges}},
  \href{https://doi.org/10.1016/S0550-3213(98)00497-0}{\emph{Nucl. Phys.}
  {\bfseries B531} (1998) 256}
  [\href{https://arxiv.org/abs/hep-th/9803004}{{\ttfamily hep-th/9803004}}].

\bibitem{Larsen:1999pp}
F.~Larsen, \emph{{Rotating Kaluza-Klein black holes}},
  \href{https://doi.org/10.1016/S0550-3213(00)00064-X}{\emph{Nucl. Phys.}
  {\bfseries B575} (2000) 211}
  [\href{https://arxiv.org/abs/hep-th/9909102}{{\ttfamily hep-th/9909102}}].

\bibitem{Giusto:2009qq}
S.~Giusto, J.~F. Morales and R.~Russo, \emph{{D1D5 microstate geometries from
  string amplitudes}},
  \href{https://doi.org/10.1007/JHEP03(2010)130}{\emph{JHEP} {\bfseries 1003}
  (2010) 130} [\href{https://arxiv.org/abs/arXiv:0912.2270}{{\ttfamily
  arXiv:0912.2270}}].

\bibitem{Bianchi:2017sds}
M.~Bianchi, D.~Consoli and J.~F. Morales, \emph{{Probing Fuzzballs with
  Particles, Waves and Strings}},
  \href{https://doi.org/10.1007/JHEP06(2018)157}{\emph{JHEP} {\bfseries 06}
  (2018) 157} [\href{https://arxiv.org/abs/1711.10287}{{\ttfamily
  1711.10287}}].

\bibitem{Witten:2000mf}
E.~Witten, \emph{{BPS Bound states of D0 - D6 and D0 - D8 systems in a B
  field}}, \href{https://doi.org/10.1088/1126-6708/2002/04/012}{\emph{JHEP}
  {\bfseries 04} (2002) 012}
  [\href{https://arxiv.org/abs/hep-th/0012054}{{\ttfamily hep-th/0012054}}].

\bibitem{Bini:2017xzy}
D.~Bini and T.~Damour, \emph{{Gravitational spin-orbit coupling in binary
  systems, post-Minkowskian approximation and effective one-body theory}},
  \href{https://doi.org/10.1103/PhysRevD.96.104038}{\emph{Phys. Rev.}
  {\bfseries D96} (2017) 104038}
  [\href{https://arxiv.org/abs/1709.00590}{{\ttfamily 1709.00590}}].

\bibitem{Vines:2017hyw}
J.~Vines, \emph{{Scattering of two spinning black holes in post-Minkowskian
  gravity, to all orders in spin, and effective-one-body mappings}},
  \href{https://doi.org/10.1088/1361-6382/aaa3a8}{\emph{Class. Quant. Grav.}
  {\bfseries 35} (2018) 084002}
  [\href{https://arxiv.org/abs/1709.06016}{{\ttfamily 1709.06016}}].

\bibitem{Bini:2018ywr}
D.~Bini and T.~Damour, \emph{{Gravitational spin-orbit coupling in binary
  systems at the second post-Minkowskian approximation}},
  \href{https://doi.org/10.1103/PhysRevD.98.044036}{\emph{Phys. Rev.}
  {\bfseries D98} (2018) 044036}
  [\href{https://arxiv.org/abs/1805.10809}{{\ttfamily 1805.10809}}].

\bibitem{Guevara:2018wpp}
A.~Guevara, A.~Ochirov and J.~Vines, \emph{{Scattering of Spinning Black Holes
  from Exponentiated Soft Factors}},
  \href{https://arxiv.org/abs/1812.06895}{{\ttfamily 1812.06895}}.

\end{thebibliography}
\end{document}